\documentclass[english,10pt,reprint,aip,jcp]{revtex4-1}
\usepackage[latin9]{inputenc}
\usepackage{xcolor}
\usepackage{booktabs}
\usepackage{bm}
\usepackage{amsmath}
\usepackage{amssymb}
\usepackage{graphicx}
\usepackage{esint}

\usepackage{amsthm}
\usepackage{graphics}
\usepackage{times}
\usepackage{bm}
\usepackage{color}
\usepackage{braket}
\usepackage{babel}

\usepackage{hyperref}

\usepackage{pdfpages}
\makeatletter
\AtBeginDocument{\let\LS@rot\@undefined}
\makeatother

\begin{document}
\title{Representing Molecular Ground and Excited Vibrational Eigenstates
with Nuclear Densities obtained from Semiclassical Initial Value Representation
Molecular Dynamics}
\author{Chiara \surname{Aieta}}
\affiliation{Dipartimento di Chimica, Università degli Studi di Milano, via C. Golgi
19, 20133 Milano, Italy}
\author{Gianluca \surname{Bertaina}}
\affiliation{Dipartimento di Chimica, Università degli Studi di Milano, via C. Golgi
19, 20133 Milano, Italy}
\affiliation{Istituto Nazionale di Ricerca Metrologica, Strada delle Cacce 91,
10135 Torino, Italy}
\author{Marco \surname{Micciarelli}}
\email{marco.miccia@gmail.com}

\affiliation{Dipartimento di Chimica, Università degli Studi di Milano, via C. Golgi
19, 20133 Milano, Italy}
\author{Michele \surname{Ceotto}}
\email{michele.ceotto@unimi.it}

\affiliation{Dipartimento di Chimica, Università degli Studi di Milano, via C. Golgi
19, 20133 Milano, Italy}
\begin{abstract}
We present in detail and validate an effective Monte Carlo approach
for the calculation of the nuclear vibrational densities via integration
of molecular eigenfunctions that we have preliminary employed to calculate
the densities of the ground and the excited OH stretch vibrational
states in protonated glycine molecule {[}C. Aieta et. al. Nat Commun
11, 4348 (2020){]}. Here, we first validate and discuss in detail
the features of the method on a benchmark water molecule. Then, we
apply it to calculate on-the-fly the ab initio anharmonic nuclear
densities in correspondence of the fundamental transitions of NH and
CH stretches in protonated glycine. We show how we can gain both qualitative
and quantitative physical insight by inspection of different one-nucleus
densities and assign a character to spectroscopic absorption peaks
using the expansion of vibrational states in terms of harmonic basis
functions. The visualization of the nuclear vibrations in a purely
quantum picture allows us to observe and quantify the effects of anharmonicity
on the molecular structure, and to exploit the effect of IR excitations
on specific bonds or functional groups, beyond the harmonic approximation.
We also calculate the quantum probability distribution of bond-lengths,
angles and dihedrals of the molecule. Notably, we observe how in the
case of one type of fundamental NH stretching the typical harmonic
nodal pattern is absent in the anharmonic distribution.
\end{abstract}
\maketitle

\section{Introduction}

Visualizing molecular vibrations in real time and at the atomistic
length scale would be of great importance to understand chemical phenomena.
Experiments can usually access molecular motions only in an indirect
way. Even if modern vibrational spectroscopies are so sophisticated
as to probe isolated molecules,\citep{fausto2019hydantoins,gerlich2018infrared,roithova2016helium,cismesia2017infrared,wolk2013cryogenic,asmis2007vibrational,quack1990spectra}
only vibrational frequencies are routinely measured. Information about
vibrational motion is extracted from the spectra via the assignment
of the peaks. In this way, for instance, it has been possible to elucidate
the structure of bio-molecules conformers,\citep{rizzo2014cryogenic}
characterize the intermediates in chemical reactions,\citep{schwarz2019identification,garand2018spectroscopy}
help the rationalization of weak interactions like H-bonds,\citep{czarnecki2015advances}
and support the understanding of solvation.\citep{voss2018accessing,asmis2011vibrational,bush2007evidence}
However, this spectroscopic procedure sometimes does not bring to
undisputed interpretations.\citep{choi2006four,gabas2018protonated}
To directly observe molecular vibrations it would be necessary to
push the limit of spatial and energy resolution of experimental spectroscopy.
A technique which evolved in this direction is the Tip-Enhanced Raman
Spectroscopy (TERS).\citep{stipe1998single,duan2016visualization,liu2017single}
Recently, a TERS experiment has produced two-dimensional spatial images
at \r{A}ngstr{\"o}m-scale resolution, where the peaks correlate
with the intensity and direction of vibrational normal-mode displacements.\citep{lee2019visualizing}
Also, data from elastic scattering of X-ray generated with a Free-Electron
Laser (XFEL) source can be opportunely treated to get diffraction
images of specific vibrational states of molecules in the gas phase.\citep{carrascosa2017imaging}

Alongside experimental techniques, a complementary way to get atomistic
insights about molecular motions is provided by computer simulations.
Theoretical chemistry methods usually simulate molecular vibrations
under the Born-Oppenheimer (BO) approximation. The dynamics of a bound
state on the BO Potential Energy Surface (PES) is usually treated
in the small oscillation regime. This approach approximates the potential
in the surroundings of a minimum as a quadratic function of the coordinates,
and the normal-mode picture for vibrations is introduced. Several
methodologies have been developed to project normal-mode motions onto
chemically meaningful internal coordinates, such as bond lengths,
angles, and dihedrals, in the most unambiguous way possible.\citep{teixeira2018improving}
However, these methods for the visualization of nuclear motion completely
overlook the effects derived from the quantum nature of the nuclei.

Solving the nuclear time-independent Schr{\"o}dinger equation from
direct diagonalization of the exact molecular Hamiltonian to get the
vibrational eigenfunction can be achieved only for low-dimensional
systems. For larger systems, it is necessary to develop specific methodologies
to account for anharmonicity and coupling between modes in the ground
and excited vibrational states.\citep{bowman1986self,gerber1988self,kosztin1996introduction,christiansen2007vibrational,christiansen2012selected,mancini_Bowman_mizedwaterHClclusters_2014,mccoy2006diffusion,bowman2008variational,matyus2009variational,ruud2000efficient,aastrand2000calculation}
Then, even if one is able to get the eigenfunctions, a further issue
is how to better visualize them and get physical information. For
N-atom molecules, the vibrational eigenfunctions have a 3N $-$ 6
dimensionality, or 3N $-$ 5 if the molecule has a linear shape. As
a consequence, such wavefunctions are usually analyzed by plotting
bidimensional cuts along some selected pairs of normal modes bringing
some information.\citep{meyer2007full} For example, the presence
of nodal planes in these contour plots and their tilted shape reveal
the resonance and anharmonic couplings between normal modes.\citep{ceotto2011firstscwf}
Visualizing directly the vibrational behavior of molecules in three-dimensional
space in the quantum picture would boost our physical insight.

Very recently, the analysis of quantum one-nucleus densities, i.e.
the probability of finding each nucleus in a molecule at a given position
in space independently of the location of the others, has been proposed
as a tool to get information about molecular normal modes from the
wavefunction.\citep{schild2019probability} In that work, harmonic
one-nucleus densities were computed by analytic integration of the
harmonic eigenfunctions. The focus was on how the wavefunction nodal
structure of the vibrationally excited states is reflected in the
one-nucleus density. It was found that only certain vibrational excitations
change the one-nucleus density qualitatively as compared to the ground
state. In alternative one can partially represent the density by including
the lighter nuclei in the electronic structure calculation.\citep{culpitt2019enhancing}

In our previous work,\citep{NatCommDens2020} we further advanced
the investigation of one-nucleus densities. More specifically, we
introduced a well-controlled Monte Carlo integration to compute expectation
values of the nuclear density operators from anharmonic vibrational
molecular eigenstates written on a basis of harmonic states. As a
first application, we estimated the expansion coefficients for the
ground and excited OH stretch eigenfunctions of protonated glycine
beyond the harmonic approximation with a semiclassical technique recently
developed in our group.\citep{micciarelli2018anharmonic} We represented
one-nucleus densities with the cube file format, that can be visualized
with 3D graphics programs,\citep{VMD} as it is customarily done for
electron density and orbitals. Comparison between the isosurface plots
of harmonic and anharmonic vibrational densities permits to visualize
molecular geometries and vibrations from a quantum mechanical point
of view.

In this work we include anharmonicity effects through the Multiple
Coherent states Time Averaged Semiclassical Initial Value Representation
(MC SCIVR),\citep{kaledin2003time,kaledin2003timeappl,ceotto2009multiple,ceotto2009first,ceotto2010multiple,ceotto2011fighting,conte2013reproducing,tamascelli2014graphics}
in which a semiclassical propagator is obtained by stationary phase
approximation of the exact Feynman\textquoteright s path integral
formulation.\citep{feynman_pathintegral_1965} Recently, Semiclassical
Initial Value Representation techniques have advanced the field of
spectroscopy simulations.\citep{liu2011insights,liu2018critical,PhysRevLett.125.083001},
including temperature-dependent spectra.\citep{micciarelli2018anharmonic,micciarelli2019effective,beguvsic2020fly}
MC SCIVR employs information obtained by processing few classical
trajectories either on the adiabatic electronic PES or on-the-fly.
Moreover, with this technique, the anharmonic excited states are obtained
at the same cost of the ground-state wavefunction. MC SCIVR had been
successfully applied for power spectra calculations, i.e. eigenvalue
calculations, to a number of medium- and large-size molecular systems
like fullerene,\citep{Ceotto2017PRL} glycine,\citep{Gabas2017Glicina}
benzene,\citep{DiLiberto2018benzene} water clusters,\citep{DiLiberto2018WaterClusters},
pre-reactive complexes,\citep{ma2018quantum}, system-bath models,\citep{Ceotto_Buchholz_SAM_2018,Ceotto_Buchholz_MixedSC_2017,Buchholz_Ceotto_MixedSC_2016}
the protonated glycine dimer and H$_{\text{2}}$-tagged protonated
glycine,\citep{gabas2018protonated} nucleobases and nucleosides,\citep{Gabas2019NucleoBasis,GabasJCTC2020},
the Zundel cation\citep{bertaina2019zundel} and more recently to
surface adsorbed molecules.\citep{cazzaniga2020anharmonic}

In this paper, we calculate the one-nucleus densities and density
differences, and discuss the shape and the nodal structure of the
corresponding excited states with an extensive application on water
molecule as a benchmark and protonated glycine, to complement the
already investigated excited OH stretch nuclear density.\citep{NatCommDens2020}
Specifically, the differences between anharmonic and harmonic wavefunctions
with the same degree of excitation highlight the effect of anharmonicity
and the relevant consequences on the molecular structure (probability
distributions of bond lengths, angles and dihedrals). In addition,
we investigate the character of vibrational excitations by considering
differences between the excited and the ground-state densities. In
the harmonic picture, we propose this as an alternative way to intuitively
visualize normal-mode displacements, without resorting to a classical
interpretation based on classical trajectories visualization. In the
anharmonic framework, these differences reveal the non-local nature
of vibrational excitations, which are directly associated to the peaks
observed in vibrational spectroscopy, going beyond the simplified
harmonic normal-mode picture.

The paper starts with the definition of the density distributions
and the description of the numerical approach we use to calculate
them. In Section \ref{sec:Theory}, we recall the semiclassical technique
based on MC SCIVR, which allows the calculation of the ground and
excited state semiclassical vibrational eigenfunctions. Next, we move
on to the presentation of the results for two representative cases
(Section \ref{sec:Results-and-Discussion}). We compute densities
for the water molecule, for which we calculate the exact values on
the same fitted PES. We use this system to validate our approach.
In Section \ref{subsec:Protonated-Glycine}, we move to the protonated
glycine (GlyH$^{+}$), a moderate dimensionality molecule containing
11 atoms for which a fitted PES is not available. This molecule is
interesting for future study of molecular solvation, as suggested
by IR spectroscopy results.\citep{marx2016waterinduced,voss2018accessing}
Finally, in the last Section, we provide the conclusions and future
development outlook.

\section{Theory\label{sec:Theory}}

\subsection{Molecular Nuclear Densities}

\textcolor{black}{Under the BO approximation, an N-atom molecule can
be described by considering the spectral decomposition of the nuclear
Hamiltonian operator $\hat{H}\ket{e_{n}}=E_{n}\ket{e_{n}}$,} where
the nuclear eigenfunctions are denoted by $\ket{e_{n}}$, and $E_{n}$
is the corresponding eigenvalue.

By diagonalizing the mass-scaled potential Hessian matrix at equilibrium
the diagonal matrix $\mathbf{\boldsymbol{\Gamma}}$ of the eigenvalues
$\omega_{\alpha}$ ($\alpha=1,\dots,3N$) is obtained, as well as
the conversion matrix between the Cartesian and the normal-mode coordinates
$\mathbf{Q}$. In this work, we analytically determine the roto-translational
modes \textbf{$\boldsymbol{Q}^{RT}$,} that we keep fixed at their
null equilibrium position, and perform a Gram-Schmidt orthogonalization
of the remaining N$_{\text{v}}$ vibrational modes.\citep{wilson1980molecular,bertaina2019zundel}
This small-displacements approximation is commonly used and corresponds
to separating vibrations and rotations, namely to neglect their coupling.
When the system lies in the n-th eigenstate $\ket{e_{n}}$, the quantum
probability density distribution of a generic physical quantity $\boldsymbol{\theta}(\boldsymbol{Q})$
is then computed as

\begin{equation}
\rho_{n,\boldsymbol{\theta}}(\boldsymbol{x})=\int d^{3N}\boldsymbol{Q}|\braket{\boldsymbol{Q}|e_{n}}|^{2}\delta(\boldsymbol{Q}^{RT})\delta(\boldsymbol{\theta}(\boldsymbol{Q})-\boldsymbol{x}),\label{eq:rho-n-theta}
\end{equation}
where $\boldsymbol{x}$ is a vector variable of the same dimensionality
as $\boldsymbol{\theta}$.

Numerically, we represent the $\rho_{n,\boldsymbol{\theta}}(\boldsymbol{x})$
density as a histogram divided into $B$ bins of volume $\Omega$
and centered around the \textbf{x}$^{j}$ $(j=1\dots B)$ positions.
Therefore, the average value of $\rho_{n,\boldsymbol{\theta}}(\boldsymbol{x})$
in the j-th bin is
\begin{equation}
\overline{\rho}_{n,\boldsymbol{\theta}}^{j}=\frac{1}{\Omega}\int d^{3N}\boldsymbol{Q}\left|\braket{\boldsymbol{Q}|e_{n}}\right|^{2}\delta(\boldsymbol{Q}^{RT})I_{\boldsymbol{\theta}}^{j}\left(\boldsymbol{Q}\right),\label{eq:rho-bar-1}
\end{equation}
where, for a given coordinate $\boldsymbol{Q}$, the index function
$I_{\theta}^{j}\left(\boldsymbol{Q}\right)$ is equal to 1 if $\boldsymbol{\theta}(\boldsymbol{Q})$
belongs to the bin centered around \textbf{x}$^{j}$, while it is
null otherwise. The density normalization condition imposes the following
normalization over its histogram representation $\Omega\sum_{j}\overline{\rho}_{n,\boldsymbol{\theta}}^{j}=1$.
The calculation of the integral in Eq. \ref{eq:rho-bar-1} is particularly
suited for its evaluation via simple Gaussian Monte Carlo sampling.
Thanks to the separation of vibrations from rotations and translations,
the eigenfunctions are expanded in the basis of the harmonic vibrational
states $\ket{\phi_{\boldsymbol{K}}}$ as

\begin{equation}
\ket{e_{n}}=\sum_{\boldsymbol{K}}C_{n,\boldsymbol{K}}\ket{\phi_{\boldsymbol{K}}}.\label{eq:en-harm-expansion-1}
\end{equation}
Here, $\boldsymbol{K}=\left(K_{1}\dots K_{N_{v}}\right)$ are positive
integer vectors, indicating the excitation degree of each harmonic
vibrational mode. The harmonic case is simply retrieved by considering
$C_{n,\boldsymbol{K}}=\delta_{\bar{\boldsymbol{K}},\boldsymbol{K}}$.
In coordinate representation, one can factor out the Gaussian terms
as $\braket{\boldsymbol{Q}|e_{n}}=G(\mathbf{Q},\boldsymbol{\Gamma})\sum_{{\bf K}}C_{n,{\bf K}}~\bar{{\bf \phi}}_{\mathbf{K}}(\mathbf{Q})$,
where $G(\mathbf{Q},\boldsymbol{\Gamma})=\left|\boldsymbol{\Gamma}/(\pi\hbar)\right|^{1/4}\exp{\left(-\mathbf{Q}{}^{T}\boldsymbol{\Gamma}\mathbf{Q}/(2\hbar)\right)}$,
and $\bar{{\bf \phi}}_{\mathbf{K}}(\mathbf{Q})=\prod_{\alpha=1}^{N_{v}}\left(2^{K_{\alpha}}K_{\alpha}!\right)^{-1/2}~h_{K_{\alpha}}\left(\sqrt{\omega_{\alpha}/\hbar}~Q_{\alpha}\right)$,
with the K$^{th}{}_{\alpha}$-order Hermite polynomial denoted as
h$_{K_{\alpha}}$. Thanks to this factorization, Eq. \ref{eq:rho-bar-1}
can be conveniently recast as
\begin{equation}
\begin{aligned}\overline{\rho}_{n,\boldsymbol{\theta}}^{j}= & \frac{1}{\Omega}\int\left[d^{3N}\boldsymbol{Q}~|G(\mathbf{Q},\boldsymbol{\Gamma})|^{2}\right]\times\\
 & \left|\sum_{{\bf K}}C_{n,{\bf K}}\bar{{\bf \phi}}_{\mathbf{K}}(\mathbf{Q})\right|^{2}\delta(\boldsymbol{Q}^{RT})I_{\boldsymbol{\theta}}^{j}\left(\boldsymbol{Q}\right).
\end{aligned}
\label{eq:trans-dip-integral-for-MC-1}
\end{equation}
Eventually, we generate a set of independent $L$ molecular configurations
along a multivariate Gaussian distribution with null mean and variance
equal to $(2\boldsymbol{\Gamma}/\hbar)^{-1}$ for the vibrational
modes only, by means of the Box-Muller algorithm,\citep{box1958muller}
and we evaluate the integral in Eq. \ref{eq:trans-dip-integral-for-MC-1}
as
\begin{equation}
\overline{\rho}_{n,\boldsymbol{\theta}}^{j}=\lim_{L\rightarrow\infty}\frac{1}{\Omega L}\sum_{l=1}^{L}\left|\sum_{{\bf K}}C_{n,{\bf K}}\bar{{\bf \phi}}_{\mathbf{K}}(\mathbf{Q}_{l})\right|^{2}I_{\boldsymbol{\theta}}^{j}\left(\boldsymbol{Q}_{l}\right).\label{eq:trans-dip-integral-with-BM-MC}
\end{equation}
Since the function $\boldsymbol{\theta}(\boldsymbol{Q})$ is always
an analytical expression of the nuclear coordinates, the computation
of Eq. \ref{eq:trans-dip-integral-for-MC-1} is computationally cheap,
once the anharmonic expansion coefficients $C_{n,{\bf K}}$ are known
for a given vibrational state $n$. Also, the values of $\bar{{\bf \phi}}_{\mathbf{K}}(\mathbf{Q}_{l})$
are analytical and can be easily evaluated over a large number of
configurations (usually in the order of $L=10^{8}$) with limited
computational overhead. By only sampling the common Gaussian term,
all samples are uncorrelated, and physical quantities relative to
multiple excited states can be sampled at once, unlike the diffusion
Monte Carlo algorithm which can measure quantities only in the ground
state or in states with predetermined nodal surface.\citep{mancini_Bowman_mizedwaterHClclusters_2014,mccoy2006diffusion}
All calculations in this work are converged in order to have statistical
errorbars that are not visible in the plots. The errorbar of the quantities
are estimated in the standard way, as the square root of the variance
divided by $L$.

When $\boldsymbol{\theta}=\boldsymbol{R}_{i}$, where $\boldsymbol{R}_{i}$
is the Cartesian position of the i-th nucleus, the probability density
of Eq. \ref{eq:rho-n-theta} assumes the form

\begin{equation}
\begin{aligned}\rho_{n,\boldsymbol{R}_{i}}(\boldsymbol{R})= & \int d^{3N}\boldsymbol{Q}\left|\braket{\boldsymbol{Q}|e_{n}}\right|^{2}\delta(\boldsymbol{Q}^{RT})\delta(\boldsymbol{R}_{i}(\boldsymbol{Q})-\boldsymbol{R})\end{aligned}
\end{equation}
which corresponds to the marginal i-th one-nucleus density,\citep{schild2019probability}
that is the nuclear analogue of electron density of Density Functional
Theory for electronic structure calculations.\citep{parr1995density}
Due to the larger mass of the nuclei as compared to electrons, the
one-nucleus densities are sufficiently localized so that the overlap
of densities of different nuclei in the molecule is negligible. Therefore,
we can consider the one-nucleus density for a molecule

\begin{equation}
\rho_{n}(\boldsymbol{R})=\sum_{i=1}^{N}\rho_{n,\boldsymbol{R}_{i}}(\boldsymbol{R})\label{eq:one-nucleus-density}
\end{equation}
which is defined in Cartesian coordinate space,\citep{schild2019probability}
and allows for the visualization of its 3D isosurfaces, as it is commonly
done for electronic structure calculations.

In this work, we also evaluate bond-length quantum distributions,
by considering $\theta=|\bm{r}_{ij}|$, where $\bm{r}_{ij}\equiv\bm{R}_{i}-\bm{R}_{j}$,
for all pairs of nuclei i, j, and angle quantum distributions, by
using $\theta=\arccos{(\hat{\bm{r}}_{ik}\cdot\hat{\bm{r}}_{jk})}$,
for all triplets of nuclei i, j, k forming an angle with vertex k,
with $\hat{\bm{r}}_{ij}\equiv\bm{r}_{ij}/|\bm{r}_{ij}|$. Finally,
we evaluate dihedral quantum distributions, for quadruplets of atoms
i, j, k, l, by considering $\theta=\text{\ensuremath{\arctan}2}{(s,c)}$,
with $s=[(\hat{\bm{r}}_{ji}\times\hat{\bm{r}}_{kj})\times(\hat{\bm{r}}_{kj}\times\hat{\bm{r}}_{lk})]\cdot\hat{\bm{r}}_{kj}$
and $c=(\hat{\bm{r}}_{ji}\times\hat{\bm{r}}_{kj})\cdot(\hat{\bm{r}}_{kj}\times\hat{\bm{r}}_{lk})$.\citep{Blondel_Newformulationderivatives_1996}

\subsection{MC-SCIVR anharmonic eigenfunctions\label{subsec:MC-TA-SCIVR-anharmonic-eigenfunc}}

\textcolor{black}{To calculate the coefficients $C_{n,{\bf K}}$ of
Eq.}\ref{eq:en-harm-expansion-1}\textcolor{black}{, we employ our
recently developed semiclassical method,\citep{micciarelli2018anharmonic,micciarelli2019effective}
which is summarized in this section.}

The eigenvectors \textcolor{black}{of a generic Hamiltonian $\hat{H}$
}are a complete basis set, and the spectroscopic weight of a given
state $\ket{\chi}$ at the energy of each eigenvalue $E_{n}$, i.e.
$\left|\braket{\chi|e_{n}}\right|^{2}$, can be obtained from the
following Fourier transform

\begin{flalign}
\tilde{I}_{\chi}(E) & =\frac{1}{\pi\hbar}Re\int_{0}^{\tau}dt\bra{\chi}e^{-\frac{i}{\hbar}\hat{H}t}\ket{\chi}e^{\frac{i}{\hbar}Et}=\nonumber \\
 & =\frac{1}{\pi\hbar}Re\int_{0}^{\tau}dt\sum_{n}\braket{\chi|e_{n}}e^{-\frac{i}{\hbar}E_{n}t}\braket{e_{n}|\chi}e^{\frac{i}{\hbar}Et}=\nonumber \\
 & =\sum_{n}\left|\braket{\chi|e_{n}}\right|^{2}\mathcal{D}(E-E_{n};\Delta_{\tau}).\label{eq:Rec-T-ovelap-power}
\end{flalign}
In the last equality, the dynamical convolution function $\mathcal{D}$
is a nascent delta function, i.e. one of the function belonging to
the sequence of functions approaching, in the weak sense, the Dirac
delta distribution, with peak centered on $E_{n}$ with amplitude
$\Delta_{\tau}$ approaching zero as the simulation time $\tau\rightarrow\infty$.
We derive the Hamiltonian eigenvalues from the positions of the spectral
peaks, while the squared projections $\left|\braket{\chi|e_{n}}\right|^{2}$
of the reference state onto the eigenvectors are determined from their
peak intensities. Specifically, the harmonic weights of Eq.\ref{eq:en-harm-expansion-1}
can be written as $\text{\ensuremath{\left|C_{n,\boldsymbol{K}}\right|}}^{2}\propto\tilde{I}_{\phi_{\boldsymbol{K}}}\left(E_{n}\right).$
As shown in detail in our previous work,\citep{micciarelli2018anharmonic}
the signed $C_{n,\boldsymbol{K}}$ coefficients can be calculated
from survival amplitudes using the following formula

\begin{align}
C_{n,\mathbf{K}}=\frac{\Delta\tilde{I}_{\phi_{\textbf{0}},\phi_{\textbf{K}}}(E_{n})}{2\sqrt{\tilde{I}_{\phi_{{\boldsymbol{0}}}}(E_{n})}},\label{eq:TA-coeff-last}
\end{align}
where $\phi_{\mathbf{0}}$ is the harmonic ground state, $\tilde{I}_{\phi_{{\boldsymbol{k}}}}(E_{n})$
is the value at energy $E_{n}$ of the power spectrum obtained with
the harmonic state $\ket{\phi_{\mathbf{K}}}$, and

\begin{align}
\Delta\tilde{I}_{\phi_{\mathbf{K}_{1}},\phi_{\mathbf{K}_{2}}}(E)\equiv & \tilde{I}_{\phi_{\mathbf{K}_{1}}+\phi_{\mathbf{K}_{2}}}(E)-\tilde{I}_{\phi_{\mathbf{K}_{1}}}(E)-\tilde{I}_{\phi_{\mathbf{K}_{2}}}(E).\label{eq:survival-harmo-Delta}
\end{align}
We obtain the quantum time evolution and the Fourier transform in
Eq.\ref{eq:Rec-T-ovelap-power} by using the MC-SCIVR approach, which
relies on the evolution of just a handful of selected classical trajectories
with initial conditions $\left(\mathbf{Q}_{0}^{(n)},\boldsymbol{P}_{0}^{(n)}\right)$.\citep{ceotto2009first,ceotto2009multiple,ceotto2010multiple,Gabas2017Glicina}
These are tailored to ideally correspond to the $n$-th vibrational
state via the Einstein-Brillouin-Keller (EBK) rules
\begin{equation}
\oint\limits _{H\left(\mathbf{Q}_{0}^{(n)},\mathbf{P}_{0}^{(n)}\right)=E_{n}}\boldsymbol{P}^{(n)}d\boldsymbol{Q}^{(n)}=\hbar\left(\zeta_{n}+\frac{\text{\ensuremath{\mu}}_{n}}{4}\right)
\end{equation}
where $\zeta_{n}$ are positive integers, and $\mu_{n}$ are Maslov
indexes.\citep{Keller_Correctedbohrsommerfeldquantum_1958} In the
separable case, these rules provide a link between the $n$-th vibrational
state and a $N_{v}$-dimensional vector of natural numbers \textbf{$\boldsymbol{\nu}$},
such that $\zeta_{n}=\sum_{\alpha}\nu_{\alpha}$, which is valid also
beyond the harmonic approximation. In the MC-SCIVR approach, these
classical trajectories are chosen with total energy (and energy partition)
corresponding to the harmonic oscillator spectral energies $E_{\pmb{\nu}}^{HO}=\sum_{\alpha}\left(1/2+\nu_{\alpha}\right)\hbar\omega_{\alpha}$,
and are generated by considering the initial conditions

\begin{flalign}
Q_{0,\alpha}^{\left(n\right)} & =\sqrt{\frac{\hbar\left(2\nu_{\alpha}+1\right)}{\omega_{\alpha}}}\sin(\delta_{\alpha})\nonumber \\
P_{0,\alpha}^{\left(n\right)} & =\sqrt{(2\nu_{\alpha}+1)\hbar\omega_{\alpha}}\cos(\delta_{\alpha}),\label{eq:MC-initial-conds}
\end{flalign}
where the equilibrium position is located at the origin and the angles
$\delta_{\alpha}$ govern the partition of the starting energy of
the $\alpha$-th normal mode into potential and kinetic terms.

In this framework, the power spectrum of the survival amplitude of
a generic state $\ket{\chi}$ is computed from the classical evolution
of a single trajectory as\citep{kaledin2003time,kaledin2003timeappl}

\begin{flalign}
\tilde{I}_{\pmb{\nu},\textbf{\ensuremath{\chi}}}(E) & \propto\dfrac{1}{\tau}\left|\int_{0}^{\tau}dt\braket{\chi|\mathbf{Q}_{t}^{(n)},\mathbf{P}_{t}^{(n)}}e^{i[S_{t}^{(n)}+\phi_{t}^{(n)}+Et]/\hbar}\right|^{2}.\label{eq:MC_power_spectrum}
\end{flalign}
When $\ket{\chi}=\ket{\phi_{\boldsymbol{K}}}$, the factor $\braket{\chi|\mathbf{Q}_{t}^{(n)},\boldsymbol{P}_{t}^{(n)}}$
is analytical, and we get
\begin{equation}
\text{\ensuremath{\left|C_{n,\boldsymbol{K}}\right|}}^{2}\propto\tilde{I}_{\phi_{\boldsymbol{K}}}\left(E_{n}\right)\simeq\tilde{I}_{\pmb{\nu},\phi_{\boldsymbol{K}}}(E_{n}).\label{eq:exp-coeff-sq-module}
\end{equation}
 In Eq. (\ref{eq:MC_power_spectrum}), the coherent states $\ket{\boldsymbol{Q},\boldsymbol{P}}$
have the following normal-mode coordinate representation\citep{Heller_FrozenGaussian_1981,Heller_SCspectroscopy_1981,Heller_Cellulardynamics_1991}
\begin{equation}
\braket{\boldsymbol{x}|\boldsymbol{Q},\boldsymbol{P}}=\left|\frac{\boldsymbol{\Gamma}}{\pi\hbar}\right|^{\frac{1}{4}}e^{-\frac{1}{2\hbar}\left(\boldsymbol{x}-\boldsymbol{Q}\right)^{T}\mathbf{\boldsymbol{\Gamma}}\left(\boldsymbol{x}-\boldsymbol{Q}\right)+\frac{i}{\hbar}\boldsymbol{P}\left(\boldsymbol{x}-\boldsymbol{Q}\right)},\label{eq:coherent-states}
\end{equation}
 $S_{t}^{\text{\ensuremath{\left(n\right)}}}$ is the classical action
of the trajectory at time t, and $\phi_{t}^{\left(n\right)}$ is the
phase of the Herman-Kluk prefactor $C_{t}^{(n)}$.\citep{Miller_Atom-Diatom_1970,Herman_Kluk_SCnonspreading_1984,Kay_Integralexpression_1994,Kay_Multidim_1994,Kay_Numerical_1994,Antipov_Nandini_Mixedqcl_2015,Church_Ananth_Filinov_2017,Wehrle_Vanicek_Oligothiophenes_2014,Wehrle_Vanicek_NH3_2015}
The latter accounts for quantum fluctuations and is defined as
\begin{equation}
C_{t}^{(n)}=\left|\frac{{1}}{2}\left(\boldsymbol{M}_{\boldsymbol{QQ}}+\boldsymbol{\mathbf{\varGamma}}^{-1}\boldsymbol{M}_{\boldsymbol{PP}}\boldsymbol{\mathbf{\varGamma}}-i\boldsymbol{M}_{\boldsymbol{QP}}\boldsymbol{\mathbf{\varGamma}}+i\boldsymbol{\varGamma}^{-1}\boldsymbol{M}_{\boldsymbol{PQ}}\right)\right|^{\frac{{1}}{2}}.\label{eq:HK-prefactor}
\end{equation}
The prefactor requires the evaluation of the stability matrix subblocks
$\boldsymbol{M}_{\boldsymbol{QQ}}=\partial\boldsymbol{Q}_{t}^{(n)}/\partial\boldsymbol{Q}_{0}^{(n)}$,
$\boldsymbol{M}_{\boldsymbol{PP}}=\partial\boldsymbol{P}_{t}^{(n)}/\partial\boldsymbol{P}_{0}^{(n)}$,
$\boldsymbol{M_{QP}}=\partial\boldsymbol{Q}_{t}^{(n)}/\partial\boldsymbol{P}_{0}^{(n)}$
and $\boldsymbol{M_{\boldsymbol{PQ}}}=\partial\boldsymbol{P}_{t}^{(n)}/\partial\boldsymbol{Q}_{0}^{(n)}$,
which are computed along each trajectory via numerical integration
of their symplectic equations of motion.\citep{brewer1997semiclassical}
For this purpose, the instantaneous Hessian matrix is needed along
each classical trajectory. This is the most computationally-expensive
part of these calculations. Specific algorithms have been developed
to reduce the computational cost in high-dimensional applications.\citep{ceotto2013evaluating,ceotto2013accelerated,Conte2019database}

\section{Results and Discussion\label{sec:Results-and-Discussion}}

\subsection{\texorpdfstring{H$_{2}$O}{H2O} Molecule\label{subsec:H2O-Molecule}}

\subsubsection{Computational details}

Some of us\citep{micciarelli2018anharmonic} recently obtained the
first 5 vibrational eigenstates of the non-rotating water molecule
using the analytical PES by Thiel et al.\citep{dressler1997anharmonic}
with the MC SCIVR method. The eigenstates were reproduced by running
five classical trajectories with initial conditions chosen according
to Eq.(12) and delta\_alpha=0, i.e. with initial momenta such that
the kinetic energy is equal to the harmonic vibrational energy of
the corresponding harmonic states (0,0,0), (0,1,0), (0,2,0), (1,0,0)
and (0,0,1). Here the spectroscopic notation reports respectively
the symmetric, bending and asymmetric normal mode quantum numbers.
The basis set was composed of the first 11 harmonic states for each
degree of freedom, implying a total of 1331 coefficients. It was shown
that a good agreement with the exact Discrete Variable Representation
(DVR) calculations can be achieved by dropping all the coefficients
smaller than 0.01 and enforcing orthonormalization by applying the
Gram-Schmidt algorithm.\citep{micciarelli2018anharmonic} In the present
work, we compute the nuclear densities from the eigenfunctions generated
with the same setup, but with a smaller threshold (equal to 10$^{-3}$)
and keeping more coefficients in the harmonic base expansion. In addition,
we have pruned the basis set by keeping only those basis functions
which have the same symmetry as the target eigenfunction. The coefficients
are reported in the Supplementary Material. For the one-nucleus densities
we used bins of edge 0.0229\r{A}, while for the bond-length distributions
we used a bin size of 0.0077\r{A} and for the angular distributions
a bin size of $0.45$ degrees. The Monte Carlo integration has been
carried out with $L=10^{8}$ steps.

\subsubsection{Anharmonicity effect on nuclear densities}

\begin{figure*}
\centering{}\includegraphics[scale=0.5]{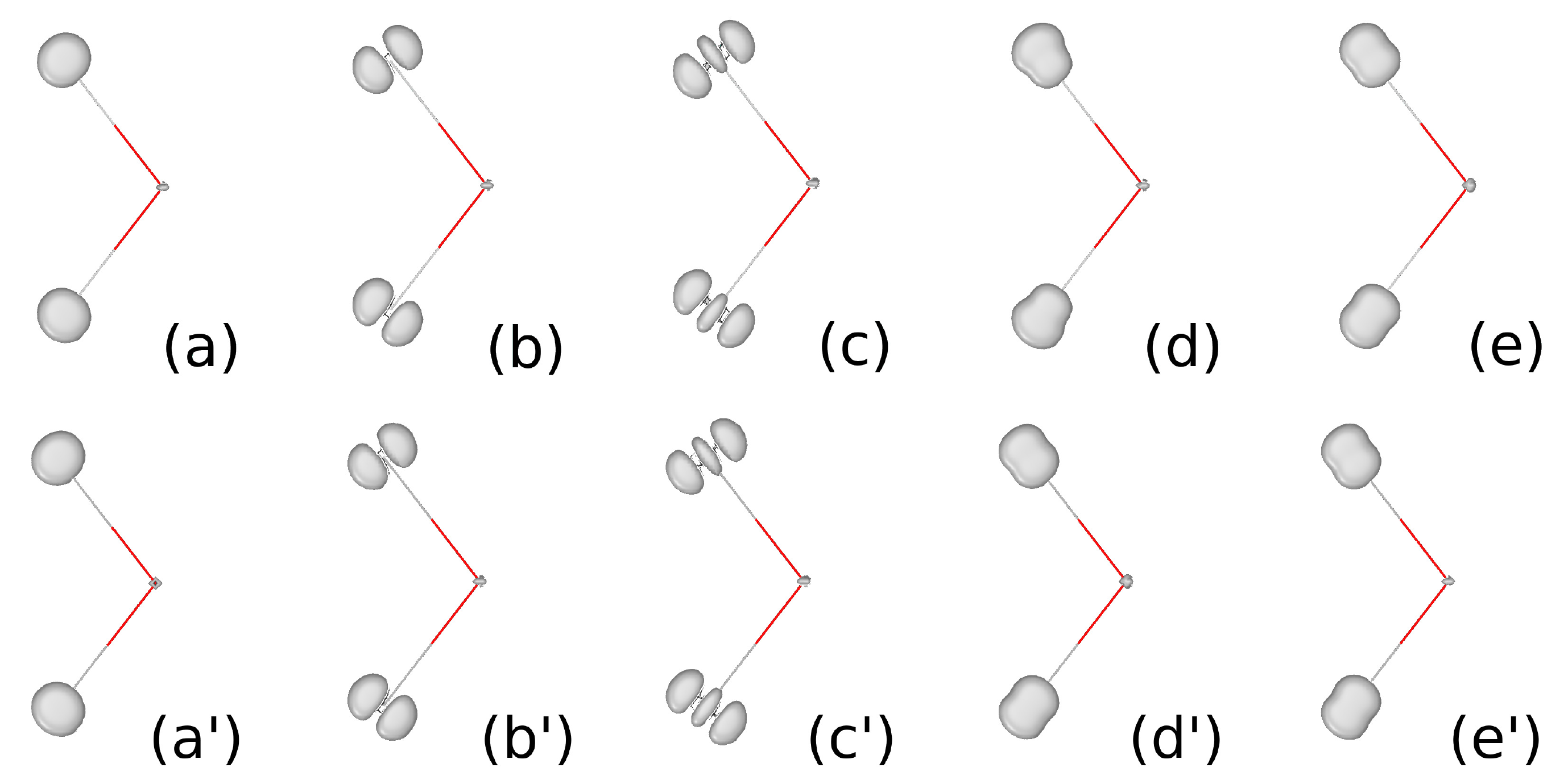}\caption{Water molecule one-nucleus densities. Anharmonic vibrational state
eigenfunction (000) in panel (a), (010) in panel (b), (020) in panel
(c), (100) in panel (d), and (001) in panel (e). Harmonic vibrational
eigenfunction (000) in panel (a'), (010) in panel (b'), (020) in panel
(c'), (100) in panel (d'), and (001) in panel (e'). All isodensity
surfaces are set to 10 a.u..\label{fig:Water-nuclear-densities}}
\end{figure*}

In Fig. \ref{fig:Water-nuclear-densities} the one-nucleus densities
of Eq. \ref{eq:one-nucleus-density} for the lowest 5 vibrational
energy eigenstates are reported. 
 For comparison, we compute also the harmonic one-nucleus densities.
The plots are shown in the water molecular plane because we do not
account for rotation. First we observe that the envelopes of the Hydrogen
densities are wider than the ones of the Oxygen. This immediately
spots the quantum nature of nuclei in molecules, whose wavefunctions
are more and more delocalized as the particle mass decreases. Then,
we observe the appearance of nodes as the quantum number increases,
however this is not guaranteed in the one-nucleus density representation.
As already observed in literature,\citep{schild2019probability} for
the harmonic case (lower panel in Fig. \ref{fig:Water-nuclear-densities}),
it is likely that the nodal structure of the wavefunction is reflected
in the one-nucleus densities when normal mode atomic displacements
are along a certain direction. In the water molecule case, the ground-state
density correctly does not show any node. The first and second excitations
of the bending mode (panels (b), (b'), (c) and (c') in Fig. \ref{fig:Water-nuclear-densities})
have respectively one and two nodal planes perpendicular to the bending
motion direction of the Hydrogens because the bending motion is only
represented by the H-O-H angle distortion. Otherwise, the first excitations
of both symmetric and asymmetric stretching equally imply a motion
along the two O-H bond distance directions. In this case the nodes
are not present, but just a depletion of one-nucleus density is observed
where one would expect the appearance of the node.

We found a similar shape of the one-nucleus anharmonic densities (upper
panels in Fig. \ref{fig:Water-nuclear-densities}). In these cases,
a deformation of the lobes appears and minor differences are visible
by direct comparison with the harmonic results. In particular a slight
tilting of the nodal planes of the bending modes is observed.

\begin{figure*}
\begin{centering}
\includegraphics[scale=0.5]{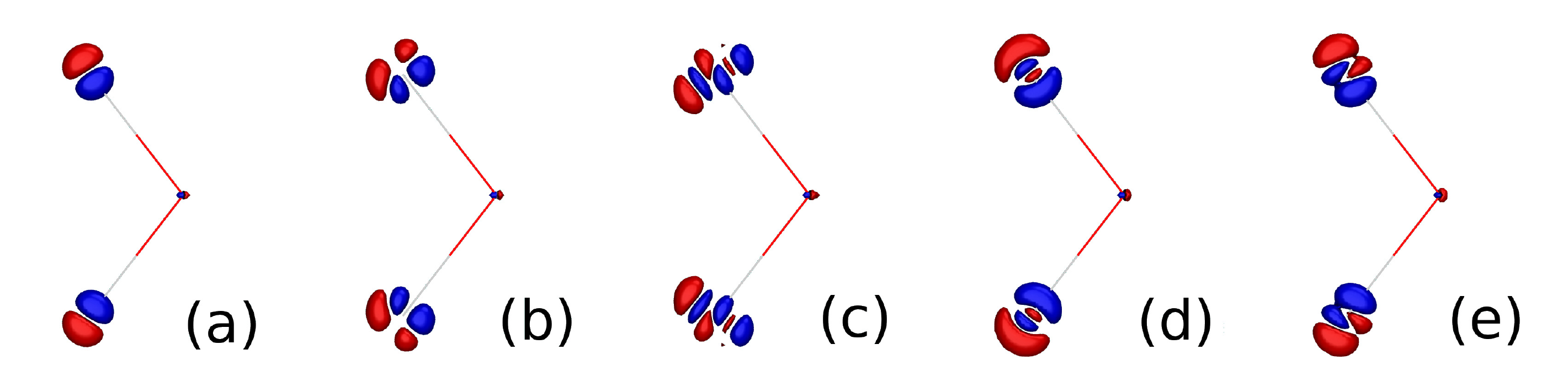}
\par\end{centering}
\caption{Differences of the anharmonic one-nucleus density with the corresponding
harmonic one. State (000) in panel (a), (010) in (b), (020) in (c),
(100) in (d), and (001) in (e). Red indicates positive contributions,
where the molecular density concentrates due to anharmonicity, while
blue stands for the negative contributions, where the density is depleted
due to anharmonicity. The isodensity surfaces are set to +5 and -5
a.u..\label{fig:H2O-Harm-Anharm-gs.}}
\end{figure*}

The difference between anharmonic and harmonic densities better clarifies
the effect of the anharmonicity, as reported in Fig. \ref{fig:H2O-Harm-Anharm-gs.}.
 For the ground state density difference (panel (a) in Fig. \ref{fig:H2O-Harm-Anharm-gs.}),
the one-nucleus density is anharmonically driven towards longer O-H
bond distances, as shown by the density accumulation (red isosurface)
and its corresponding density depletion (blue isosurface). Similar
effects are observed for all the investigated excited states. This
shows that in the anharmonic case the equilibrium distances should
be bigger than the harmonic one. In addition, for the two bending
modes (panels (b) and (c)), a smaller bond angle is expected, since
the bigger red lobes are localized in the inner part of the H-O-H
angle. As regarding the symmetric and asymmetric stretches (panels
(d) and (e)), the density differences hint at a slighter deformation
towards larger H-O-H angles for the asymmetric stretch only.

\begin{figure}
\begin{centering}
\includegraphics[width=0.9\columnwidth]{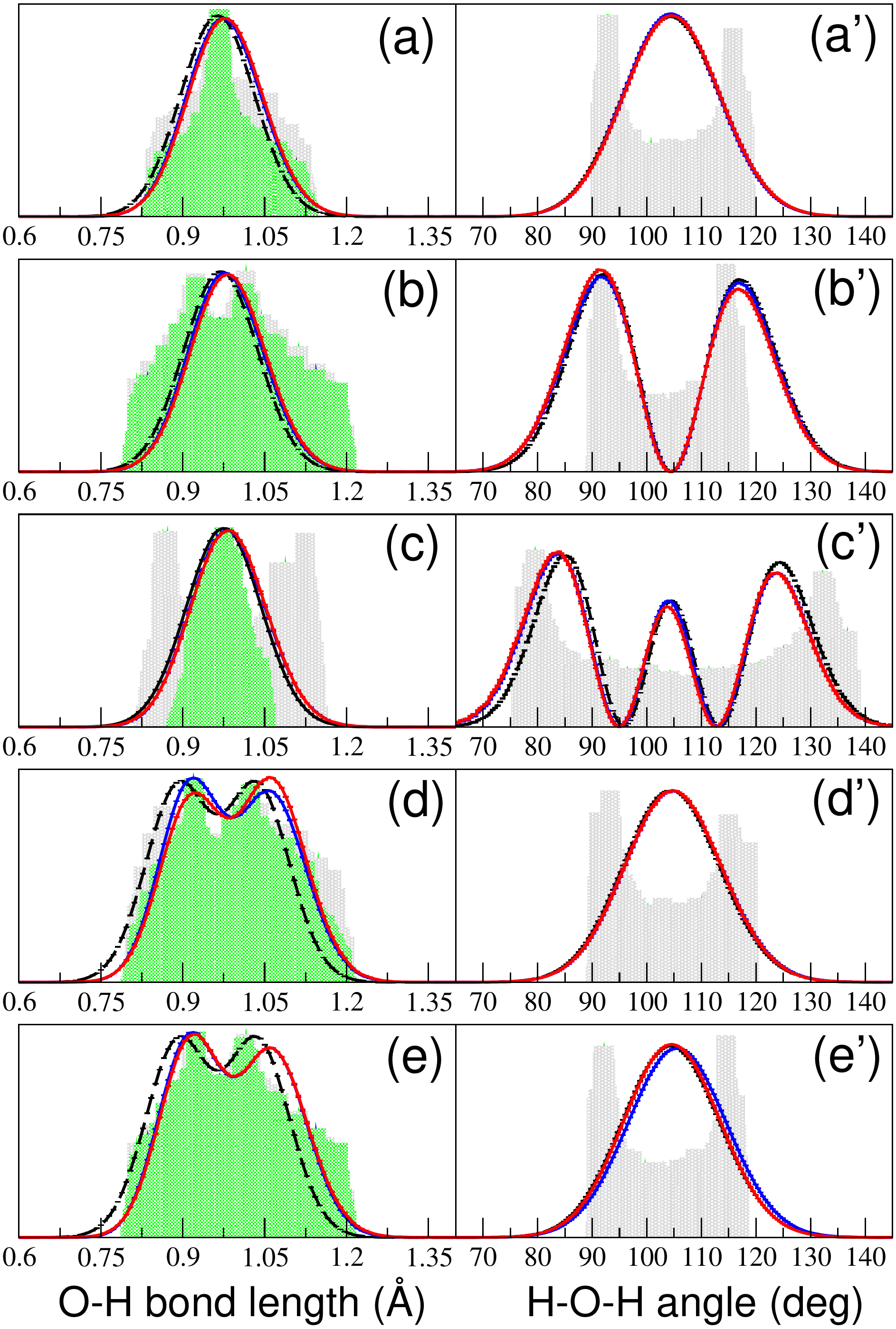}
\par\end{centering}
\caption{Water bonds and H-O-H angle probability density distributions obtained
from quantum harmonic (black dashed line), DVR \citep{micciarelli2018anharmonic}
(blue line), and semiclassical (red line) wavefunctions. Left panels
(a), (b), (c), (d), and (e) are respectively for the radial O-H distribution
of the ground (000), (010), (020), (100) and (001) vibrational eigenfunctions.
Right panels (a'), (b'), (c'), (d'), and (e') are for the angular
H-O-H distributions of the same ordered vibrational states. The filled-area
plots on the background represent the classical distributions obtained
from the classical trajectory used to generate each wavefunction with
our semiclassical approach. In the left panels, gray and green histograms
stand for the two bond-length distributions along each water O-H bond
and, in the right panels, gray histograms are for the H-O-H angle.
\label{fig:H2O-bonds-angles}}
\end{figure}

These qualitative observations, driven by visual inspection of density
differences, are confirmed by a quantitative analysis of probability
distributions of bond distances and angles amplitudes derived from
the quantum harmonic, the semiclassical and the exact quantum eigenfunctions,
the latter obtained by normal-mode DVR simulations.\citep{micciarelli2018anharmonic}
These calculations are reported in Fig. \ref{fig:H2O-bonds-angles},
where just one of the two bonding distances is plotted in the left
column, because of symmetry. 
 The statistical error bars for the distributions in Fig. \ref{fig:H2O-bonds-angles}
are smaller than the line width. In Fig. \ref{fig:H2O-bonds-angles}
the semiclassical (red line) distributions are in good agreement with
the DVR ones (blue lines). For the ground state, it is found that
the average bond lengths are slightly increased for the anharmonic
wavefunctions (panel (a)), while the bond angle is practically unaltered
(panel (a')). Moreover, for all the excited states, all bond lengths
are longer in the anharmonic picture with respect to the harmonic
one (panels (b), (c), (d) and (e)). In particular, the increase in
the bond length is more significant for the two stretching modes (panels
(d) and (e)). The anharmonicity effect of O-H bonds elongation is
consistent with the Morse-like shape of the potential along the direction
of the bonds. Conversely, the angle manifests a contraction only for
the excited bending modes (panels (b') and (c')).

In the same figure, the bond length and angle distributions derived
from the classical trajectories employed for the semiclassical simulation
are reported as either gray or green histograms. One would expect
the corresponding distribution maximum at the classical turning points
if the motion along the angle or a bond corresponds exactly to the
displacement of a single normal mode. This is clearly seen for all
the classical H-O-H angle distributions (right panels in Fig. \ref{fig:H2O-bonds-angles}),
because the angle deformation can be described by the bending mode
variation only. Notice that, for all the employed EBK trajectories,
we assign to the bending mode a kinetic energy corresponding at least
to the harmonic zero-point energy (ZPE). In contrast, the quantum
ground state and stretching excited state angle distributions are
peaked around their equilibrium positions (see panels (a'), (d') and
(e')). Instead, for the excited bending states (panels (b') and (c')),
the quantum mechanical distribution becomes more similar to the classical
one, as the quantum number is increased. A more complicated picture
arises in the bond-length distribution case (left panels of Fig. \ref{fig:H2O-bonds-angles}).
The probability distributions are equal for both O-H bonds due to
the symmetry of the wavefunction. However, this is not always the
case for the classical distributions (green and grey histograms, see
panel (c)). This happens because our classical trajectories are short-time
trajectories, as requested from the semiclassical approach, and they
are too short to guarantee equilibration of the energy between the
degrees of freedom of the molecule. Nevertheless, the comparison of
classical distributions derived from these short trajectories is still
useful to make it evident that we are able to reproduce correct quantum
mechanical results starting from classical information. Indeed, the
quantum distributions are always wider than the classical ones for
all the considered quantities. This reveals that we are actually reproducing
quantum effects, because semiclassical distribution probabilities
are non-zero in classical forbidden regions of the motion.

\subsubsection{Anharmonicity effects on vibrational excitations}

\begin{figure*}
\begin{centering}
\includegraphics[scale=0.5]{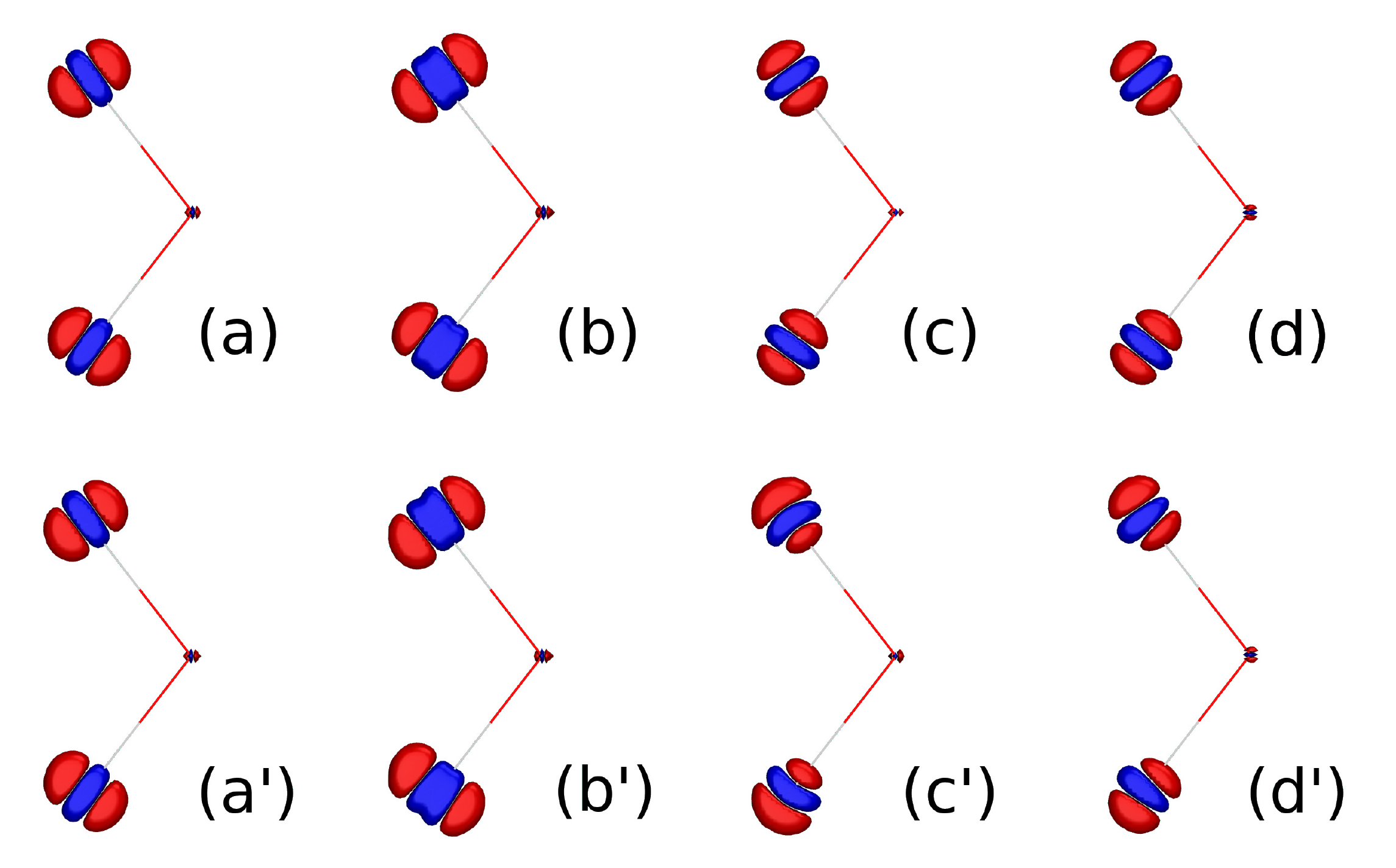}
\par\end{centering}
\caption{Harmonic (top panels (a), (b), (c), and (d)) and semiclassical anharmonic
(lower panels (a'), (b'), (c'), and (d')) nuclear density differences
obtained by subtracting the zero point energy (000) state one-nucleus
density to respectively the (010), (020), (100) and (001) excited
state densities. Red indicates positive contributions, where the density
concentrates due to the excitation, while blue stands for the negative
contributions, where the density is depleted due to the excitation.
The isodensity surfaces are set to +5 and -5 a.u..\label{fig:H2O-excitations}}
\end{figure*}

To gain deeper quantum insights into the vibrational excitations,
we propose in Fig. \ref{fig:H2O-excitations} the differences between
the excited state vibrational one-nucleus densities and the ground-state
ones. In the upper panel we report the differences in harmonic approximation,
while in the lower panel the anharmonic ones. Red lobes indicate density
concentration, while blue lobes indicate density depletion as a consequence
of the vibrational transition from the ground to the excited state.
In Fig. \ref{fig:H2O-excitations} the harmonic bending excitations
cause a deformation of the angle (panels (a), (b)), while the stretching
excitations deform the density along the two O-H bond directions (see
panels (c) and (d)). Here, one can see how the nuclear delocalization
is wider in the overtone bending excitation (panel (b)) than in the
fundamental one (panel (a)), as expected by comparison with a simple
one-dimensional harmonic oscillator. This is true because the mode
variation is given by a single angular variation. When the normal
mode involves several geometric parameters, the nuclear density variation
is not necessarily so intuitive. Asymmetric and symmetric stretching
density variation looks very similar, because the correlations between
nuclear motions are lost in this one-nucleus density picture. However,
the small deformation on the Oxygen density with vertical nodal planes
which appears in panel (c), is consistent with the symmetric stretching
motion, where the Oxygen motion keeps the molecular center of mass
fixed. In turn, for the asymmetric stretch in panel (d) the lobes
on the Oxygen have the nodal planes set horizontally.

The lower panels of Fig. \ref{fig:H2O-excitations} show the anharmonic
density variation after excitations. Also in this case the shape of
the lobes is influenced by the motion of the corresponding normal
mode that brings the biggest contribution in the harmonic expansion
of the wavefunction. Interestingly, many features show a certain amount
of anharmonicity. For example, the symmetric stretch state density
difference (panel (c')) is accumulated more towards the tip of the
two O-H bonds, due to significant anharmonicity. In the same fashion,
for the asymmetric stretch state density difference (panel (d')) the
lobes are distributed along a curved line. Finally, the fundamental
and overtone bending excitation cases (panels (a') and (b')) are more
similar to the harmonic densities, given the slightly wider lobes,
with respect to the anharmonic case.

\subsection{Protonated Glycine\label{subsec:Protonated-Glycine}}

\subsubsection{Computational Details}

\begin{figure}
\centering{}\includegraphics[scale=0.25]{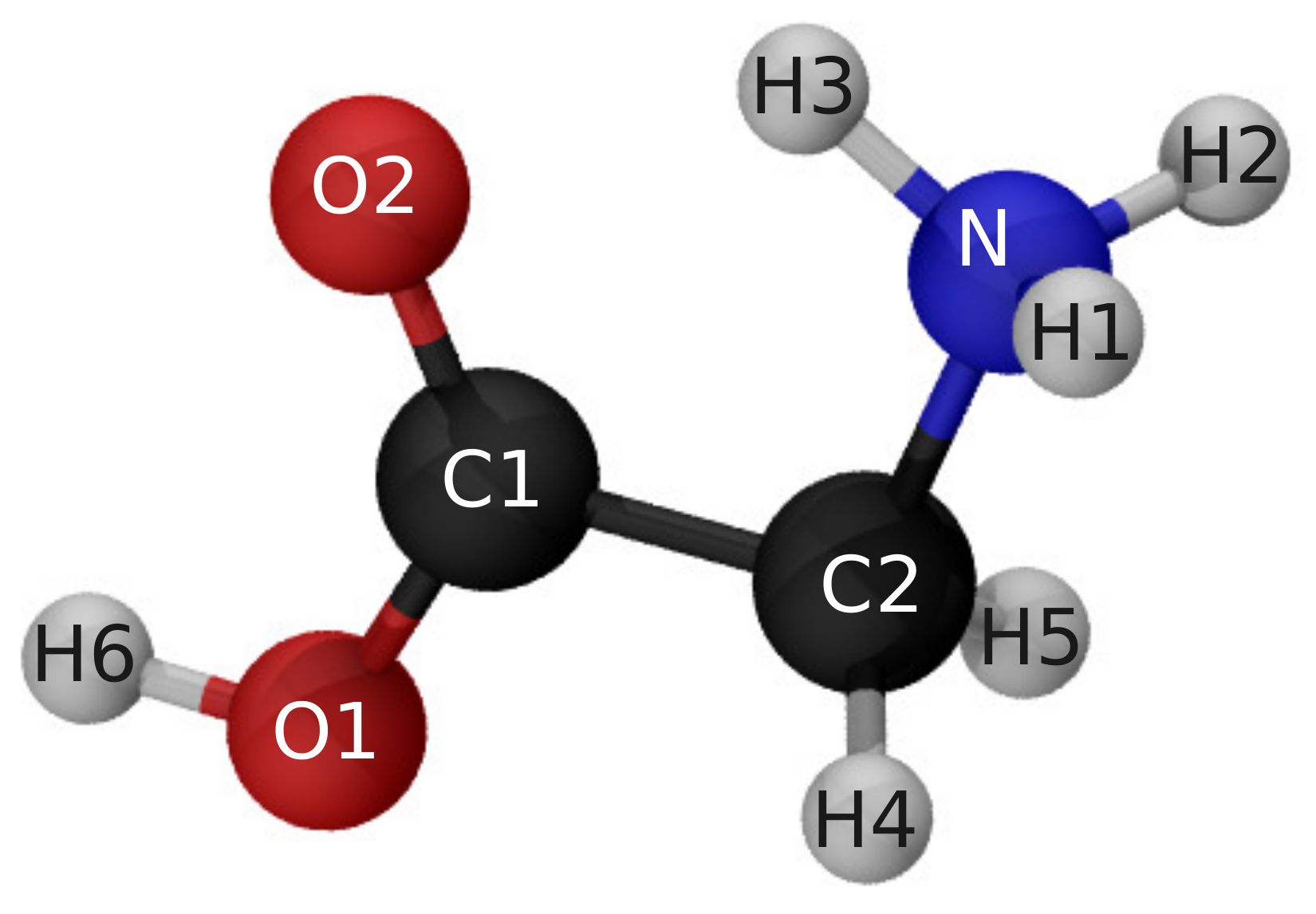}\caption{Protonated Glycine (GlyH$^{+}$) lowest energy conformer at DFT-B3LYP/aug-cc-pVDZ
level of theory.\label{fig:GlyH-geom}}
\end{figure}

In this Section we consider the 11-atom protonated Glycine (GlyH$^{+}$)
molecule. We are able to tackle such a system because the MC-SCIVR
technique can be applied on-the-fly when fitted PESs are not available.
As in our previous work,\citep{NatCommDens2020} we perform the quantum
chemistry calculations at the DFT-B3LYP level of theory using the
aug-cc-pVDZ basis set with the NWChem package.\citep{NWCHEM} The
gas phase global minimum has the protonated amino group which establishes
an ionic hydrogen bond with the carbonyl Oxygen.\citep{zhang1998conformers}The
optimized structure in Fig. \ref{fig:GlyH-geom} displays C$_{s}$
symmetry (see file ``reference\_geometry\_Glyp.xyz'' included in
the Fortran software package \citep{DensPack_2020} for Cartesian
coordinates), whose relevant normal-mode frequencies and symmetry
characters are reported in Table \ref{tab:GlyH-normal-modes}.

\begin{table*}
\centering{}\caption{The 6 highest fundamental frequencies ($\omega_{\alpha}$) for the
protonated Glycine (GlyH+) at DFT-B3LYP/aug-cc-pVDZ level of theory
together with their symmetry group Irriducible Representation (Irr.
Repr.) and their Double Harmonic Approximation (DHA) IR intensities.
\label{tab:GlyH-normal-modes}}
\begin{tabular}{ccccc}
\toprule 
$\alpha$-th normal mode & $\omega_{\alpha}$(cm$^{-1}$) & Description & Irr. Repr. & DHA IR intensity (a.u.)\tabularnewline
\midrule
22 & 3105 & N-H3 stretch + C2-H4/5 symmetric stretch in phase & A' & 0.804\tabularnewline
23 & 3117 & N-H3 stretch + C2-H4/5 symmetric stretch out of phase & A' & 4.580\tabularnewline
24 & 3170 & C2-H4/5 asymmetric stretch & A'' & 0.123\tabularnewline
25 & 3445 & N-H1/2 symmetric stretch & A' & 2.882\tabularnewline
26 & 3505 & N-H1/2 asymmetric stretch & A'' & 2.718\tabularnewline
27 & 3693 & O1-H6 stretch & A' & 4.263\tabularnewline
\bottomrule
\end{tabular}
\end{table*}

\begin{figure*}
\begin{centering}
\includegraphics[scale=0.5]{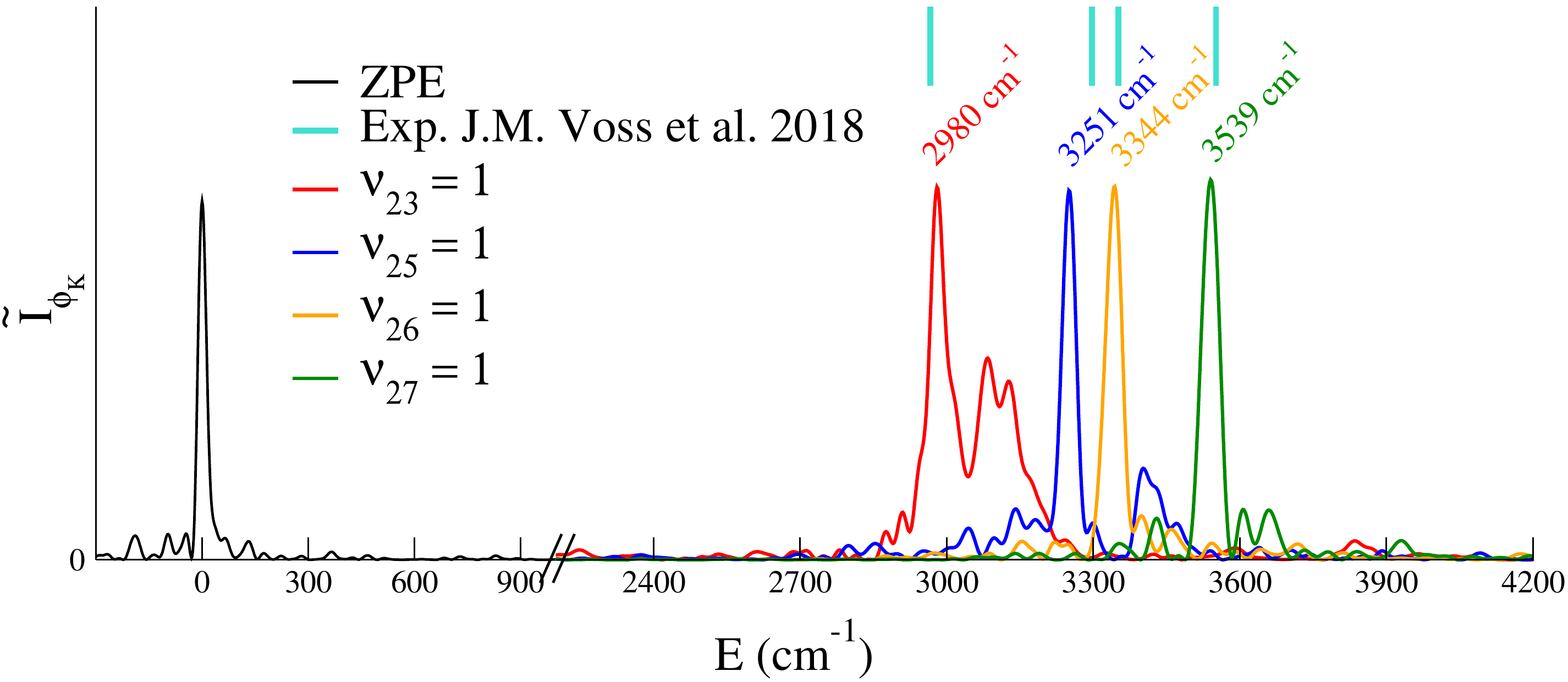}
\par\end{centering}
\caption{GlyH$^{+}$ MC-SCIVR power spectrum. Each peak is obtained from a
trajectory with energy equal to the fundamental mode 23, 25, 26, and
27 using the initial conditions of Eq. \ref{eq:MC-initial-conds}
and the relative harmonic reference state.\citep{ceotto2011fighting}
The peak positions are shown relative to the ZPE one, and the peak
heights have been arbitrarily scaled, since we are interested on peak
positions only, i.e. $E_{n}$ in Eq. \ref{eq:exp-coeff-sq-module}.
The ZPE and v$_{27}$ peaks are taken form our previous study.\citep{NatCommDens2020}
Vertical lines are the peak positions from the experimental IR spectrum\citet{voss2018accessing}.\label{fig:GlyH-power-spectrum.}}
\end{figure*}

We focus on the 2600-3800 cm$^{-1}$ region of the vibrational spectrum,
which has been investigated experimentally to get structural information
of solvated GlyH$^{+}$clusters by comparison with IR spectra of the
isolated-molecule.\citep{voss2018accessing} Specifically, we focus
on modes 23, 25, and 26, which are the IR active ones in this region,
as confirmed by the Double Harmonic Approximation (DHA) in Table \ref{tab:GlyH-normal-modes}.
The OH stretch mode 27 is also active and we have already discussed
it in our previous work.\citep{NatCommDens2020} We then run 4 on-the-fly
trajectories, each one with initial conditions corresponding to the
harmonic EBK prescription of the ZPE and the 3 fundamentals, as described
in Eq. \ref{eq:MC-initial-conds}. Given the freedom in choosing the
angle $\delta_{\alpha}$ in Eq. \ref{eq:MC-initial-conds}, we set
it equal to $\pi/2$ for normal modes 24 and 26, which correspond
respectively to the N-H1/2 and C2-H4/5 asymmetric stretches. The standard
choice of $\delta_\alpha =0$, i.e. the one for the equilibrium
position, would have required a longer simulation time for observing
the stretching of both bonds. Instead, with this choice we can better
explore the stretching motions during the short-time semiclassical
dynamics and obtain the power spectrum displayed in Fig. \ref{fig:GlyH-power-spectrum.}.
In the same figure, a part from the fundamentals, we can observe side
peaks which we attribute to the combination of each fundamental with
the low frequency modes. However, in this work our analysis is focused
on the fundamental signals. 

We write the eigenfunctions as a combination of 12799 coefficients,
after restricting the harmonic basis to the simultaneous excitation
of two modes at most, and with at most harmonic quantum number equal
to 6. The shape of the density is always determined by the largest
coefficients in the harmonic expansion. By gradually dropping the
smaller coefficients we found that just those greater than $10^{-3}$
are significant. As in the case of water, we enforce $C_{s}$ symmetry
to the harmonic-basis wavefunctions. We also apply Gram-Schmidt orthonormalization
between eigenfunctions starting from the ground state one. We report
the largest coefficients of the states analyzed in this work and the
ground state one in the Supplementary Material. As already discussed
in our previous work,\citep{NatCommDens2020} the largest expansion
coefficient in the ground-state function is the one of the harmonic
ground state (Table S2 of the Supplementary Material). Similarly,
for the wavefunctions computed in this work, we find that the largest
coefficients are those of the harmonic state with one quantum of energy
intake for mode 23 and 26 respectively (Table S3 of the Supplementary
Material). As for the $\nu_{25}=1$ anharmonic eigenfunction, there
are three leading terms in the expansion whose coefficients are comparable
(Table S4 of the Supplementary Material). However, the harmonic state
with one quantum of energy on mode 25 is the only fundamental excitation
among these three harmonic states, thus determining the overall character
of the wavefunction. The other two largest coefficients are combination
of low frequency mode harmonic eigenfunctions.

Eventually, we represent the one-nucleus density with a histogram
of 3D cubes with edge equal to 0.049 \r{A}. The Monte Carlo integration
has been carried out with L=10$^{8}$ steps. We also obtained all
the bond-length, angle, and dihedral densities with both harmonic
and anharmonic wavefunctions for all the considered states, with typical
resolutions of 0.008 \r{A}, 0.45 degrees, and 0.45 degrees, respectively.

\subsubsection{Anharmonicity effect on nuclear densities}

\begin{figure}
\centering{}\includegraphics[width=0.6\columnwidth]{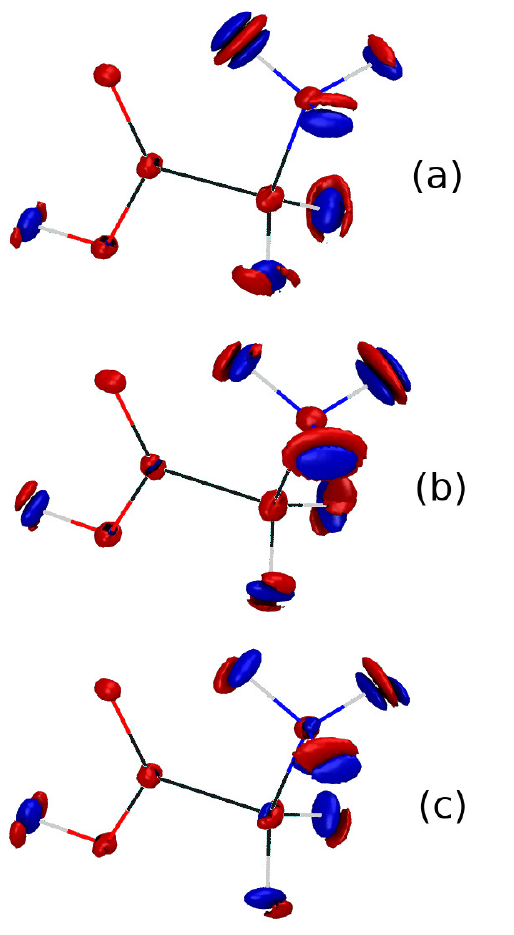}\caption{One-nucleus density differences between the anharmonic and the corresponding
harmonic marginal one-nucleus density for the (a) $\nu_{23}=1$, (b)
$\nu_{25}=1$, and (c) $\nu_{26}=1$ vibrational excited states. Red
indicates positive contributions, where the density concentrates due
to anharmonicity, while blue stands for the negative contributions,
where the density is depleted due to anharmonicity. The isodensity
surfaces are respectively set to +0.15 and -0.15 a.u..\label{fig:ZPEh-ZPE_23h-23_25h-25_26h-26}}
\end{figure}

In this paragraph we analyze, with the aid of the calculated one-nucleus
densities, bond, angle, and dihedral distributions and the effect
of the inclusion of anharmonicity in the excited vibrational states
of GlyH$^{+}$. For the sake of brevity, we only show some selected
bond-length distributions in the main text. Additional distributions
are reported in the Supplementary Material, when useful to support
the discussion.

\begin{figure}
\begin{centering}
\includegraphics[width=0.95\columnwidth]{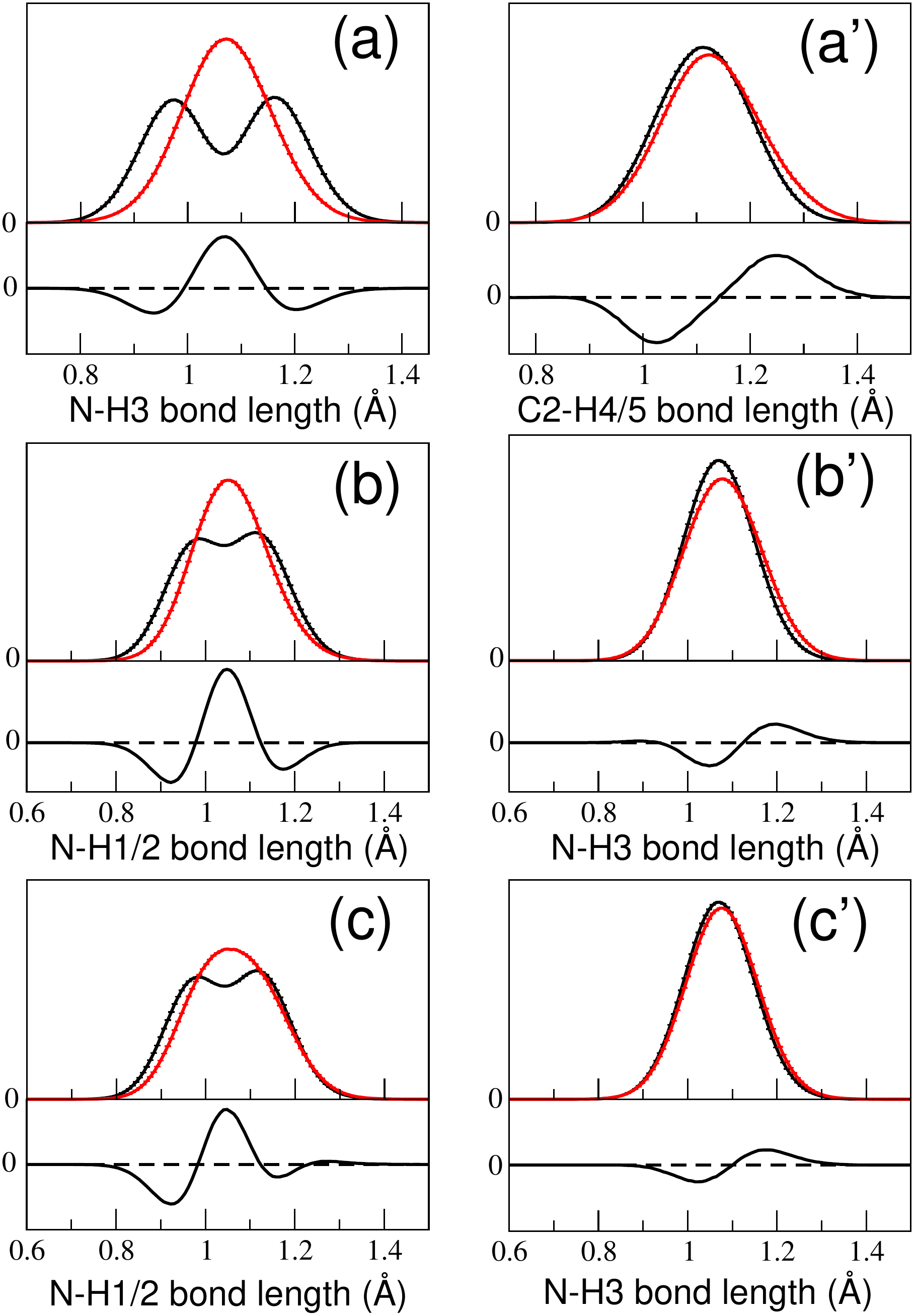}
\par\end{centering}
\caption{Bond-length distributions comparison between harmonic (black lines)
and anharmonic (red lines) states for excitation $\nu_{23}=1$ along
the N-H3 and C2-H4/5 stretching (panels (a) and (a')), $\nu_{25}=1$
along N-H1/2 and N-H3 ones (panels (b) and (b')), and $\nu_{26}=1$
along N-H1/2 and N-H3 directions (panels (c) and (c')). The lower
part of each plot shows the difference of the two curves reported
in the upper part by subtracting the harmonic distribution from the
anharmonic one.\label{fig:N-H-and-C-H-and-N-Hsh-bond-length}}
\end{figure}

We start from the excited vibrational state $\nu_{23}=1$, where the
inclusion of anharmonicity shows a less mobile H3. This localization
is evident from the isodensity plot in panel (a) of Fig. \ref{fig:ZPEh-ZPE_23h-23_25h-25_26h-26},
where the anharmonic accumulation of the H3 one-nucleus density toward
the center of the distribution is spotted by a red lobe with two symmetric
blue lobes at the side. We better render this feature by looking at
the N-H3 and C2-H4/5 bond-length distributions reported in panels
(a) and (a') of Fig. \ref{fig:N-H-and-C-H-and-N-Hsh-bond-length}.
The C2-H4/5 bond length is increased with respect to the ground state
and to the harmonic case and the H4-C2-H5 angle is larger (see Figure
S1 in the Supplementary Information). A strong anharmonic effect is
observed in panel (a) of Fig. \ref{fig:N-H-and-C-H-and-N-Hsh-bond-length}
for the N-H3 bond. Here, the harmonic double peak distribution becomes
a single peak upon anharmonicity inclusion. In the harmonic case,
the normal mode 23 has a significant displacement vector lying along
the N-H3 bond direction. This movement counterbalances the symmetric
C2-H4/5 stretch to keep the center of mass of the molecule fixed.
We can model the oscillation along the N-H3 as a one-dimensional harmonic
oscillator, and we observe the appearance of a node in the probability
distribution by giving one quantum of excitation, consistently with
the one-dimensional harmonic oscillator model. Instead, in the anharmonic
picture the oscillation along the N-H3 bond is no longer separable
from other motions. Anharmonicity mixes normal mode 23 with other
mode contributions and the associated displacements of the N and H3
nuclei no longer lie along the bond direction. As a consequence, no
clear nodal feature is found upon excitation.

For both the $\nu_{25}=1$ and $\nu_{26}=1$ excited states (panels
(b) and (c) of Fig. \ref{fig:ZPEh-ZPE_23h-23_25h-25_26h-26}), the
most evident anharmonic effect is the localization of the one-nucleus
density on the H1/2 nuclei. This apparently contradictory feature
is explained by the appearance of a single peak distribution for the
N-H1/2 bond lengths in the anharmonic picture, as shown in panels
(b) and (c) of Fig. \ref{fig:N-H-and-C-H-and-N-Hsh-bond-length}.
The N-H1/2 symmetric and asymmetric stretches can be compared to the
water symmetric and asymmetric stretches described above. From the
normal-mode point of view they behave similarly because normal modes
25 and 26 involve exclusively two H atoms distance variations without
any angle changes. However, the inclusion of anharmonicity acts differently
for the amino group. For water, we observe that the harmonic and anharmonic
bond-length distributions have the same double peak shape (panels
(d) and (e) in Fig. \ref{fig:H2O-bonds-angles}), while for GlyH$^{+}$
the distributions become single-peaked in the anharmonic case (panels
(a), (b), and (c) in Fig. \ref{fig:N-H-and-C-H-and-N-Hsh-bond-length}),
showing once again that both $\nu_{25}=1$ and $\nu_{26}=1$ excitation
dynamics involve several atoms, and not only the N-H1/2 and N-H3 distances.
Therefore, the inclusion of anharmonicity in the eigenfunction highlights
the couplings between these N-H stretches and other modes. In particular,
we find for both modes a significant coupling to the breathing of
the O2-C1-C2-N-H3 ring of atoms. Concerning mode 25, there is a general
broadening of bond length and angles distributions for the ring structure.
At the same time, a clear reduction of the O2-C1-C2 and C1-C2-N angles
is observed, while the N-H3 bond becomes longer (Fig.\ref{fig:N-H-and-C-H-and-N-Hsh-bond-length},
panel (b')). However, the asymmetric stretch state 26 predicts a deformation
of the ring structure, by shortening the O2-H3 distance and elongating
the C1-C2 backbone bond. At the same time, we observe that the C2-N
bond becomes shorter and the N-H3 longer (Fig.\ref{fig:N-H-and-C-H-and-N-Hsh-bond-length},
panel (c')). All these considerations suggest that the introduction
of anharmonicity for both states leads to a picture where the H3 is
increasingly shared with the O2 atom of the carbonyl group. Finally,
we note that the broadening of the dihedral angles distributions are
complementary in the two states. More specifically, in the $\nu_{25}=1$
eigenstate, the dihedrals distributions for the ``carboxylic end''
of the molecule are equal to the harmonic ones, while the dihedral
distributions for the ``aminic end'' result slightly broadened.
Instead, for the $\nu_{26}=1$ eigenstate, the effect is the opposite,
since anharmonicity introduces a significant broadening of the distributions
for the carboxyl part of GlyH$^{+}$. All these information are well
summarized by the isodensity plots in Fig. \ref{fig:ZPEh-ZPE_23h-23_25h-25_26h-26}.

\subsubsection{Anharmonicity effect on vibrational excitations}

We now look at the one-nucleus density differences between excited
vibrational states and the ground one, both using the harmonic and
anharmonic eigenfunctions. We will show that a significant wavefunction
spreading under excitations over the molecular structure occurs for
all the considered anharmonic eigenstates, at variance with the harmonic
ones.

\begin{figure}[btp]
\begin{centering}
\includegraphics[width=0.95\columnwidth]{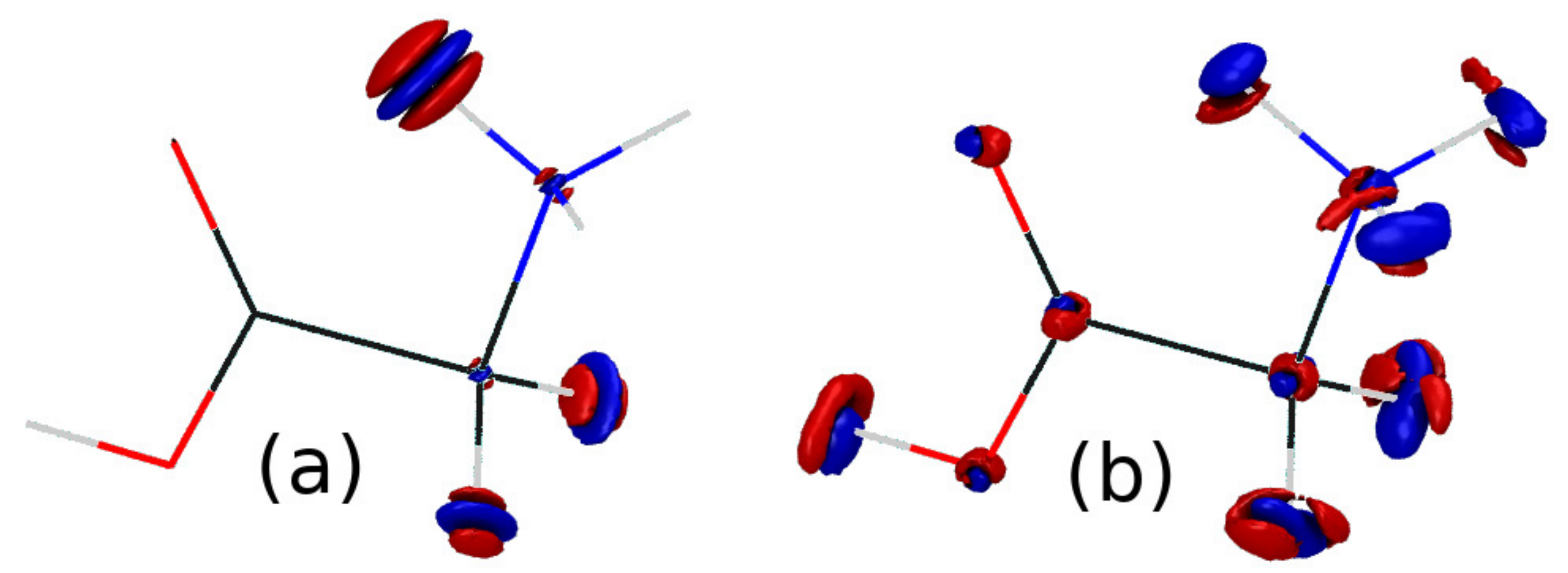}
\par\end{centering}
\centering{}\caption{GlyH$^{+}$ one-nucleus density difference plots for $\nu_{23}=1$
vibrational eigenstate. Panel (a) depicts the difference between excited
and ground harmonic densities, while panel (b) shows the anharmonic
case. Red indicates positive contributions, while blue stands for
the negative contributions. All isodensities are set to +0.1 and -0.1
a.u..\label{fig:23h-ZPEh_23-ZPE}}
\end{figure}

\begin{figure}[tb]
\begin{centering}
\includegraphics[width=0.95\columnwidth]{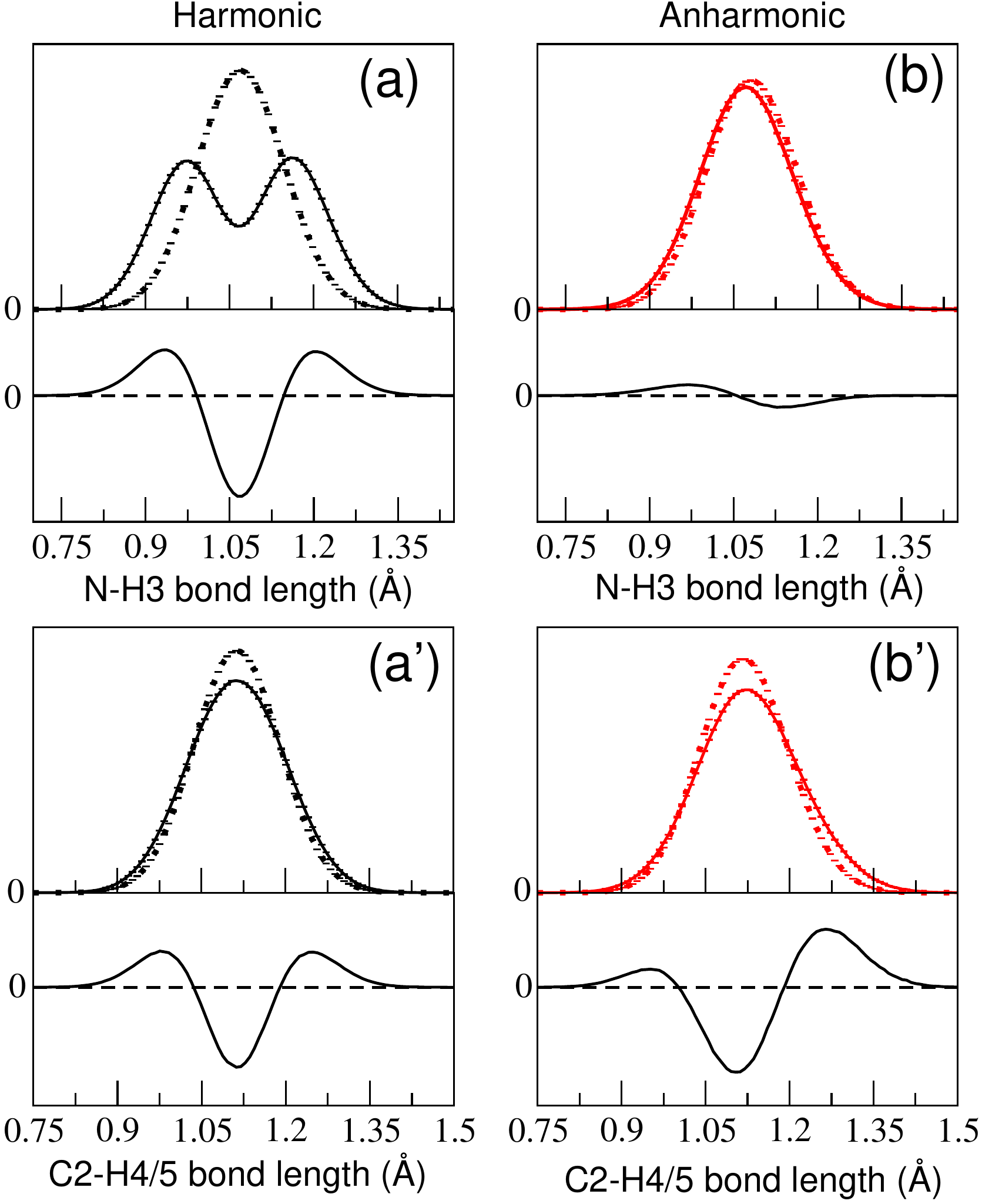}
\par\end{centering}
\caption{The $\nu_{23}=1$ excited state (continuous lines) and the ground
state (dotted lines) bond-length distributions. Panels (a) and (a')
in harmonic approximation along the N-H3 and C2-H4/5 stretch directions
respectively. Panels (b) and (b') show the anharmonic case. The ground
state densities are presented in our previous work. \citep{NatCommDens2020}
The lower parts of the plots show the difference of the curves reported
in the upper parts.\label{fig:State-23-NH-bonds-dist}}
\end{figure}

Panel (a) of Fig. \ref{fig:23h-ZPEh_23-ZPE} and panels (a) and (a')
of Fig. \ref{fig:State-23-NH-bonds-dist} show the one-nucleus density
differences in harmonic approximation for the case $\nu_{23}=1$.

 The harmonic isodensity difference plot shows lobes on the H3 nucleous,
meaning that this vibrational mode has its major contribution on the
N-H3 stretching and a minor one on the two C1-H4/5 stretches as shown
in panel (a) of Fig. \ref{fig:23h-ZPEh_23-ZPE}. In Fig. \ref{fig:State-23-NH-bonds-dist},
in the harmonic case (left panels), the N-H3 bond-length distribution
is a single peak around equilibrium, while it becomes a double peak
in the excited state $\nu_{23}=1$ (panel (a)). On the contrary, the
C2-H4/5 bond-length distribution is single-peaked both in the ground
and in the excited state (panel (a')). This agrees with the harmonic
normal-mode picture, where the bigger displacements in Cartesian coordinates
are found along the N-H3 bond direction. 

The semiclassical anharmonic eigenfunctions show instead quite a different
pattern and the harmonic node is not present in the anharmonic excitation.
As before, this is because the excitation involves several modes and
the reasoning based on the harmonic mode excitation is not meaningful
anymore. This is really apparent by inspection of Fig. \ref{fig:23h-ZPEh_23-ZPE}.
In the harmonic case, the density deformation is confined to the displacements
related to normal mode 23, while in the semiclassical anharmonic picture
the one-nucleus density change is distributed all over the molecular
structure. One can appreciate this by looking at the red lobes along
the backbone structure in the anharmonic excitation in Fig. \ref{fig:23h-ZPEh_23-ZPE}.
In the semiclassical anharmonic picture, the $\nu_{23}=1$ excitation
includes a density change for the far-away O1-H6 stretch displacement
as well. The two-lobe shape on H3 in this picture can be better understood
by looking at bond-length distributions in Fig. \ref{fig:State-23-NH-bonds-dist}.
Panel (b) reports a single-peak distribution for the N-H3 bond length
also for the excited state, differently from the harmonic double peak
one. The three-lobe shape on H4/5 is instead qualitatively equivalent
to the harmonic one, as it is observed for the C2-H4/5 bond-length
distributions (panel (a') and panel (b')). This evidence proves a
strong coupling between the oscillation along N-H3 bond and the other
modes in the anharmonic picture, and a weaker coupling of the C2-H4/5
stretches with the other motions.

\begin{figure}[bt]
\begin{centering}
\includegraphics{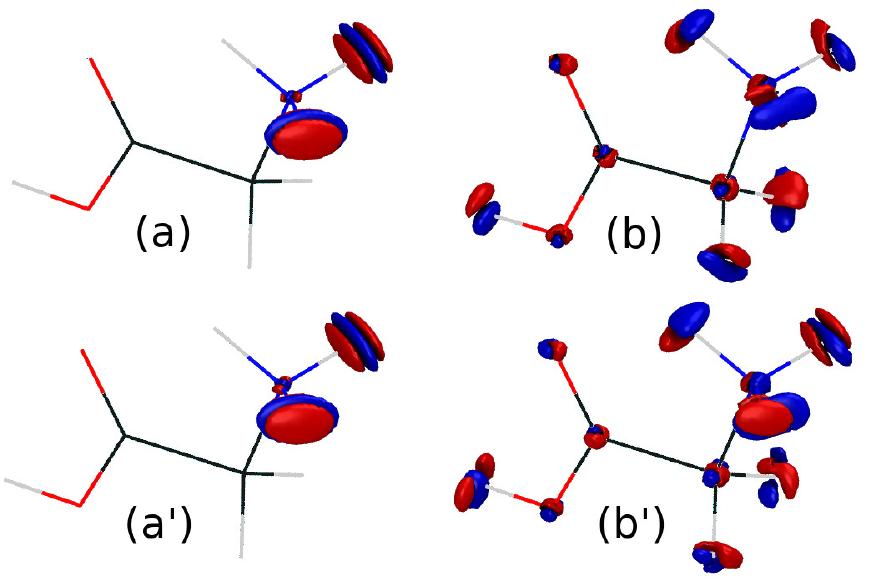}
\par\end{centering}
\caption{One-nucleus density difference plots for the GlyH$^{+}$ $\nu_{25}=1$
(panels (a) and (b)) and $\nu_{26}=1$ (panels (a') and (b')) vibrational
eigenstates. Panels (a) and (a') depict the difference of the harmonic
excited state and ground state densities, while panels (b) and (b')
refer to the anharmonic case. Red indicates positive contributions,
while blue stands for the negative contributions. All isodensities
are set to 0.1 and -0.1 a.u..\label{fig:25-26-isodens}}
\end{figure}

Moving to higher frequency modes, the density differences in harmonic
approximation indicates a very neat effect on the N-H1/2 bond oscillations
for the $\nu_{25}=1$ and $\nu_{26}=1$ states, as reported in Fig.
\ref{fig:25-26-isodens}, panels (a) and (a'). This is consistent
with the classical normal-mode displacements. In the same pictures,
the symmetric and asymmetric N-H1/2 stretches could be distinguished
only by looking at the small contributions on the N nucleus where
the distortion in the N one-nucleus density in the symmetric stretch
case is parallel to the main distortion found on the H1/2 nuclei (panel
(a)). The reason is that, as already pointed out in the case of the
water molecule, the molecule has to keep its center of mass fixed
while undergoing a symmetric stretch. A perpendicular distortion is
present in the N-H1/2 asymmetric stretch (panel (a')). Turning our
attention to the anharmonic excitations (Fig. \ref{fig:25-26-isodens},
panel (b) and panel (b')), the spreading of the excitations over the
whole structure is once again the main feature. The excitation of
the eigenstate $\nu_{25}=1$ has a less evident stretching character
than the $\nu_{26}=1$. In other words, the anharmonic lobe pattern
of the H1 and H2 atoms of the amino group is more similar to the harmonic
ones for the $\nu_{26}=1$ eigenfunction than for the $\nu_{25}=1$
state. Nevertheless, the two states can be distinguished using the
same reasoning applied to the harmonic excitations. Specifically,
the nodal planes of the two H nuclei in the amino group are set perpendicularly
to the bond axes in the case of the symmetric stretch, while for the
asymmetric stretch, they are set parallel. Another feature of the
anharmonic excitations is the shortening of O2-H3 distances, which
is more pronounced for the $\nu_{25}=1$ eigenstate. This is indicated
by the position of the two anharmonic density accumulations on the
O2 and H3 nuclei, and it is also confirmed by the corresponding bond-length
distribution. In the harmonic picture the O2-H3 distance reduction
is completely missed. There, the mode related to this intermolecular
rearrangement is the N-H3 stretch mode 23, which instead loses this
local character in the anharmonic picture, as discussed in the previous
paragraph.

\section{Summary and Conclusions}

In this paper we have detailed how semiclassical eigenfunctions,\citep{micciarelli2018anharmonic,micciarelli2019effective}
which are written as combinations of products of one-dimensional harmonic
vibrational eigenfunctions, can be profitably employed for nuclear
density calculations. Specifically, we calculate the marginal one-nucleus
density, and the bond-length, angle and dihedral distributions, using
a Monte Carlo integration over the remaining nuclei positions. Given
the state of the art where one-nucleus densities are given in harmonic
approximation,\citep{schild2019probability} our semiclassical approach
calculates the nuclear densities including anharmonic and quantum
mechanical effects. The method is based on classical trajectories
and it is implemented either on a pre-computed PES or on-the-fly,
i.e. using an ab initio molecular dynamics approach.\citep{conte2020sensitivity}

We take the water molecule as a benchmark for checking the accuracy
of our nuclear densities and get familiar with this quantum mechanical
nuclear representation. We observe the quantum harmonic description
to be quite accurate in this case and similar to the semiclassical
and exact anharmonic ones, by reproducing all main quantum features.

We then calculate protonated glycine molecule nuclear densities, complementing
our previous work.\citep{NatCommDens2020} In this case, no PES is
available and our semiclassical calculations are performed by running
classical trajectories on-the-fly, using the NWChem suite of codes.
We find, in this case, the coupling to be strong and the picture provided
by the harmonic approximation of the normal modes to be oversimplified,
since excitations are typically spread all over the molecular structure.
For example, vibrational excitations $\nu_{25}=1$ and $\nu_{26}=1$
in a normal-mode picture are described only by vibrational excitations
of the amino group, i.e. N-H1/2 stretching displacements. Instead,
in the quantum mechanical picture provided by our semiclassical nuclear
densities, all atoms are significantly affected and we find even a
strong involvement of the O1-H6 stretch, which is located at the other
end of the molecule.

In the case of water molecule, we also point out that a classical
density distribution obtained using the same trajectories employed
for the semiclassical simulations, is inadequate for the low quantum
number vibrational state density description and it becomes more suitable
as the number of quanta of excitation are increased.

Finally, we find that, when considering significantly anharmonic states,
the three-dimensional (3D) one-nucleus densities are usefully complemented
by the distributions of internal coordinates, such as bond lengths,
angles, and dihedrals, because they focus more on local distortions.
We expect that the methodology presented here will provide insightful
information also for more flexible molecules, especially when considering
the densities pertaining to rigid modes, provided a judicious sampling
of the floppy modes will be enacted.\citep{bertaina2019zundel}

In conclusion, the quantum mechanical tool presented in this paper
allows us to show and quantify both for ground and vibrational excited
states how much nuclear densities, and nuclear motion in general,
deviate from a harmonic description.

\section*{SUPPLEMENTARY MATERIAL}

See Supplementary Material, for the list of coefficients of the water
and protonated glycine vibrational wavefunctions, and for plots of
additional distributions of the protonated glycine molecule, useful
to support the discussion. We also provide a Fortran software package
\citep{DensPack_2020} with instructions for the reproduction of the
results for both water and GlyH+.

\section*{Acknowledgments}

The authors thank Dr. F. Gabas, for contribution in the early stages
of this work. The authors acknowledge financial support from the European
Research Council (ERC) under the European Union\textquoteright s Horizon
2020 research and innovation programme {[}Grant Agreement No. (647107)---SEMICOMPLEX---ERC-2014-CoG{]}
and from the Italian Ministry of Education, University, and Research
(MIUR) (FARE programme R16KN7XBRB project QURE). Part of the needed
CPU time was provided by CINECA (Italian Supercomputing Center) under
ISCRAB project ``QUASP'' and ISCRAC project ``MCSCMD''.

\section*{AIP PUBLISHING DATA SHARING POLICY}

The data that support the findings of this study are openly available
in Zenodo at http://doi.org/10.5281/zenodo.4046872, reference number
\citep{DensPack_2020}.

\bibliographystyle{aipnum4-1}
%

\onecolumngrid
\newpage

\includepdf[pages={1},width=\paperwidth]{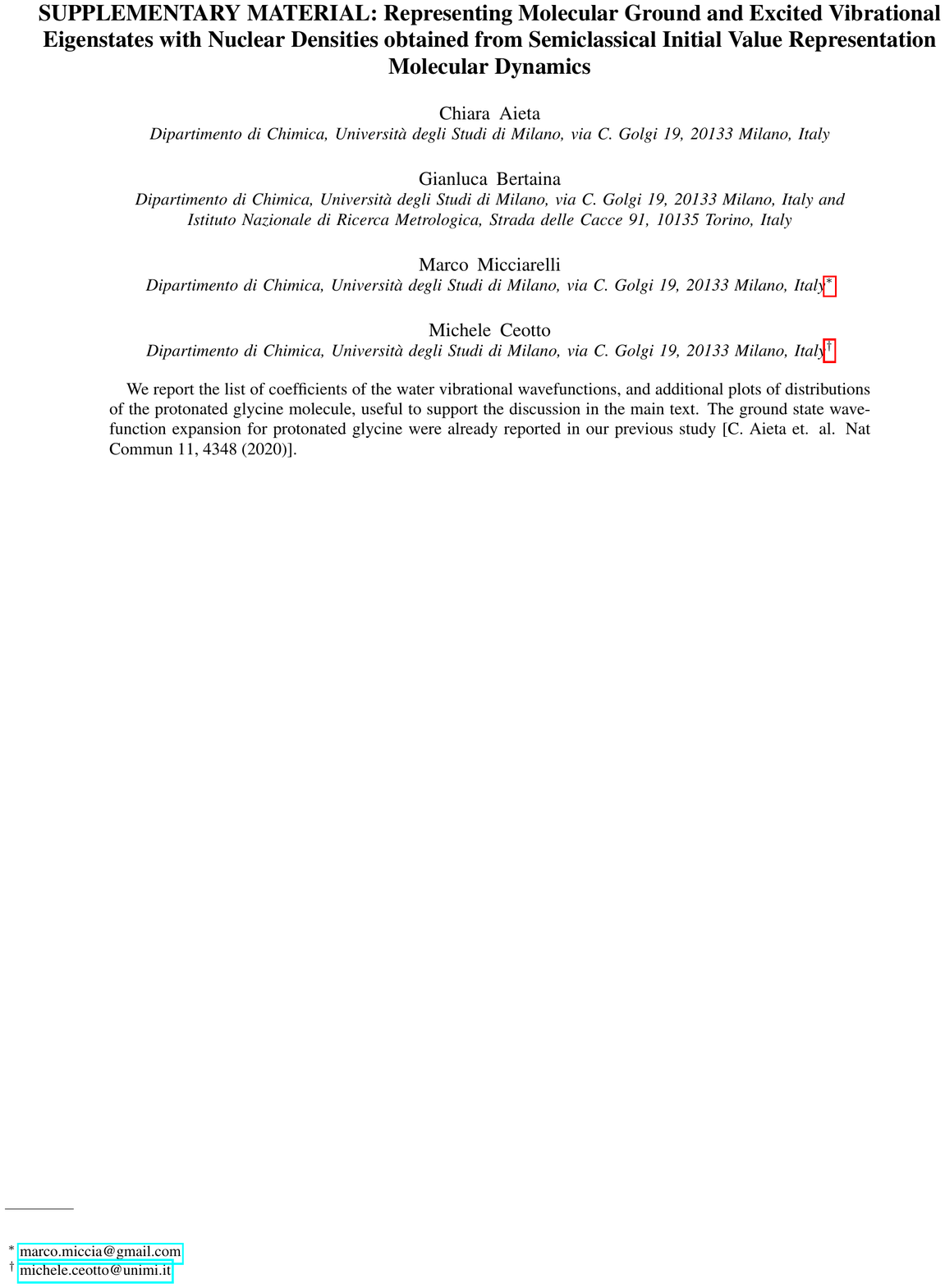}
\includepdf[pages={2},width=\paperwidth]{./Supplementary.pdf}
\includepdf[pages={3},width=\paperwidth]{./Supplementary.pdf}
\includepdf[pages={4},width=\paperwidth]{./Supplementary.pdf}
\includepdf[pages={5},width=\paperwidth]{./Supplementary.pdf}
\includepdf[pages={6},width=\paperwidth]{./Supplementary.pdf}
\includepdf[pages={7},width=\paperwidth]{./Supplementary.pdf}

\end{document}